\documentclass[twocolumn,aps,prb,showpacs,amsmath,superscriptaddress,longbibliography,notitlepage]{revtex4-2}
\usepackage{wasysym}
\usepackage{amssymb,mathtools}
\usepackage{physics}
\usepackage{mathrsfs}
\usepackage{graphicx}
\usepackage{float}
\usepackage[caption=false]{subfig}
\usepackage{dcolumn}
\usepackage{appendix}
\usepackage{tabularx}
\usepackage{epstopdf}
\usepackage[normalem]{ulem}
\usepackage{verbatim}
\usepackage{bm}
\usepackage{natbib}
\usepackage{multirow}
\usepackage{nicematrix,tikz}
\usetikzlibrary{fit}
\usepackage{xcolor}
\usepackage{hyperref}
\hypersetup{
	colorlinks,
	citecolor=cyan,
	linkcolor=red,
	urlcolor=magenta}

\DeclareMathAlphabet\mathbfcal{OMS}{cmsy}{b}{n}

\usepackage{soul}
\begin{document}
\title{Impedance responses and size-dependent resonances in topolectrical circuits \\ via the method of images}
\author{Haydar Sahin}
\email{sahinhaydar@u.nus.edu}
\affiliation{Department of Electrical and Computer Engineering, National University of Singapore, Singapore 117583, Republic of Singapore}
\affiliation{Institute of High Performance Computing, Agency for Science, Technology and Research (A*STAR), Singapore 138632, Republic of Singapore}
\author{Zhuo Bin Siu}
\email{elesiuz@nus.edu.sg}
\affiliation{Department of Electrical and Computer Engineering, National University of Singapore, Singapore 117583, Republic of Singapore}
\author{S. M. Rafi-Ul-Islam}
\email{elesmr@nus.edu.sg}
\affiliation{Department of Electrical and Computer Engineering, National University of Singapore, Singapore 117583, Republic of Singapore}
\author{Jian\nobreak \space Feng Kong}
\email{kong\_jian\_feng@ihpc.a-star.edu.sg}
\affiliation{Institute of High Performance Computing, Agency for Science, Technology and Research (A*STAR), Singapore 138632, Republic of Singapore}
\author{Mansoor B.A. Jalil}
\email{elembaj@nus.edu.sg}
\affiliation{Department of Electrical and Computer Engineering, National University of Singapore, Singapore 117583, Republic of Singapore}
\author{Ching Hua Lee}
\email{phylch@nus.edu.sg}
\affiliation{Department of Physics, National University of Singapore,Singapore 117542, Republic of Singapore}
\begin{abstract}
Resonances in an electric circuit occur when capacitive and inductive components are present together. Such resonances appear in admittance measurements depending on the circuit's parameters and the driving AC frequency. In this study, we analyze the impedance characteristics of nontrivial topolectrical circuits such as one- and two-dimensional Su–Schrieffer–Heeger circuits and reveal that size-dependent anomalous impedance resonances inevitably arise in finite $LC$ circuits. Through the \textit{method of images}, we study how resonance modes in a multi-dimensional circuit array can be nontrivially modified by the reflection and interference of current from the structure and boundaries of the lattice. We derive analytic expressions for the impedance across two corner nodes of various lattice networks with homogeneous and heterogeneous circuit elements. We also derive the irregular dependency of the impedance resonance on the lattice size, and provide integral and dimensionally-reduced expressions for the impedance in three dimensions and above.
\end{abstract}
\maketitle
\section{Introduction}
Electric circuit networks are extremely versatile platforms for simulating a variety of condensed matter phenomena through their tight-binding representations~\cite{ningyuan_time-_2015,song_realization_2020,nakata_circuit_2012}. For instance, a uniform tiling of $LC$ oscillators composed of capacitor and inductor pairs gives rise to an AC signal propagating along an ideal transmission line, which simulates the lattice dynamics of a one-dimensional (1D) solid-state medium. Circuits that mimic condensed matter lattices, known as topolectrical (TE) circuits, have been extremely successful in demonstrating a wide range of topological and critical condensed matter phenomena~\cite{lee_topolectrical_2018,imhof_topolectrical-circuit_2018,helbig_band_2019,bao_topoelectrical_2019,rafi-ul-islam_topoelectrical_2020,rafi-ul-islam_realization_2020,wang_circuit_2020,wang_circuit_2020,li_emergence_2019,ezawa_electric_2019,lee_imaging_2020,helbig_generalized_2020,zhang_topolectrical-circuit_2020,olekhno_topological_2020, ni_robust_2020,hofmann_chiral_2019,rafi-ul-islam_system_2022,shang_2022_experimental}. In these circuits, topologically protected zero modes can be measured as impedance resonances at the resonant frequency. Beyond the simulation of linear condensed matter systems, the simulation of complex networks processes such as search algorithms has also been proposed through the use of non-linear circuit elements.~\cite{ezawa_electric_2020,ezawa_universal_2021,pan_electric-circuit_2021,quiroz-juarez_reconfigurable_2021,kotwal_active_2021,kengne_ginzburglandau_2022}. %Indeed, many physical phenomena can give rise to impedance resonances in their corresponding electrical circuit arrays. 

While tight-binding lattice models are usually faithful in representing their respective solid state systems~\cite{kane_z_2005,jin_solutions_2009,sun_nearly_2011,hasan_colloquium_2010,lee_lattice_2014,chen_impossibility_2014,gu_holographic_2016,rafi-ul-islam_unconventional_2022}, their intrinsic discreteness sometimes leads to additional anomalous contributions to the impedance with no analog in the continuum limit. In this work, we present an analytic formulation for computing the impedance across various finite circuit arrays. We also investigate the anomalous impedance behavior that emerges when components with different phase lags are simultaneously present in a finite electric circuit, which we call a heterogeneous circuit. We focus most specifically on heterogeneous circuits in which the nodes are coupled by a mixture of capacitors and inductors.

The electric potential distribution in electric circuit networks satisfies Kirchhoff's laws and can be described by the circuit Laplacian. The solutions of the circuit Laplacian contain all available information about the potential distribution. Therefore, any desired computation can be performed by solving the circuit Laplacian with appropriate boundary conditions. Although the two-point impedance in homogeneous circuits has been widely studied, most results pertaining to the condensed matter context are valid only for infinite circuit networks~\cite{kirkpatrick_percolation_1973,lavatelli_resistive_1972,bartis_lets_1967,zemanian_classical_1984,cserti_perturbation_2002,owaidat_two-point_2018,owaidat_perturbation_2014,koutschan_lattice_2013,izmailian_generalised_2014,izmailian_two-point_2014,tan_resistance_2017,cserti_uniform_2011,venezian_resistance_1994,owaidat_interstitial_2010,owaidat_resistance_2019,atkinson_infinite_1999,doyle_random_2000,tan_electrical_2020,tan_two-point_2016,asad_infinite_2005,essam_exact_2009,izmailian_asymptotic_2010,clerc_dielectric_1996,cserti_application_2000,morita_useful_1971,asad_perturbed_2014,owaidat_resistance_2016,owaidat_perturbation_2016,tan_electrical_2019,tan_recursion-transform_2015,zhang_resistance_2021,chen_equivalent_2021,tan_characteristic_2017,aitchison_resistance_1964,joyce_exact_2002,jeng_random_2000,chen_electrical_2019,chen_electrical_2020,chen_electrical_2020-1,fang_circuit_2022,cernanova_nonsymmetric_2014,joyce_exact_2017,asad_infinite_2014,asad_infinite_2013,guseinov_unified_2007,mamode_calculation_2019,zenine_lattice_2015,tan_recursion-transform_2015_1,tan_recursion-transform_2015_2,owaidat_resistance_2014,owaidat_electrical_2013,owaidat_substitutional_2010,jafarizadeh_calculating_2007,giordano_disordered_2005}. Two issues that present challenges in forming full analogies with condensed matter are the finite circuit boundaries (which differ from the usual open boundary conditions) and the homogeneity of the lattice array; while computations can certainly be performed numerically, a comprehensive analytical expression for the impedance in heterogeneous finite circuits has yet to be obtained. One way to implement boundary conditions in electrostatic theory is through the method of images, in which the boundaries are replaced by image charges located opposite the original charges~\cite{jackson_classical_1999,griffiths_introduction_2005,riley_mathematical_1999,yang_designing_2022,mamode_electrical_2017}. As a demonstration of this approach, we apply the method of images to periodic tilings of finite electric circuit networks to compute the two-point impedance of the finite circuit networks. 

While the impedance generally scales in a logarithmic manner with the circuit size in homogeneous circuits \cite{wu_theory_2004,tan_recursion-transform_2015,owaidat_regular_2014,cserti_uniform_2011,cserti_perturbation_2002,owaidat_resistance_2013,asad_infinite_2014,owaidat_substitutional_2010}, the same is not true in heterogeneous circuits, where the impedance can deviate very strongly from logarithmic scaling~\cite{abrahams_scaling_1979} at certain circuit sizes. The circuit size $N$ thus becomes a functional parameter alongside $LC$ and the driving AC frequency $\omega$. These independent parameters collectively affect the impedance resonances. This is in contrast to a waveguide or transmission line, in which the system size does not affect the behavior of the system in ideal cases. Moreover, the discreteness of $N$ results in fractal-like resonances when $N$ is varied at fixed $L$ and $C$ values.
\\
\\
\indent In this work, we study the size-dependent impedance resonances both numerically and analytically. We derive analytic expressions for the size-dependent impedance between two opposite corner nodes by utilizing the method of images. To reveal the origin of this anomalous impedance behavior, we examine circuits with a single node per unit cell, as well as those with nontrivial unit cells containing more than one node. As paradigmatic examples of the latter, we present detailed calculations for 1D and two-dimensional (2D) circuit lattices with Su-Schrieffer-Heeger (SSH) type dimerizations. As for homogeneous circuits with a single-type node per unit cell, we start from a 1D circuit and build higher-dimensional circuits by linking every node along the new direction with the same type of component (i.e., resistor, inductor, or capacitor). In the heterogeneous circuits section, we follow the same process but introduce at least two different types of components with different phase shifts. Finally, we discuss the emergent fractal-like structures that arise from the violated logarithmic scaling in $LC$ circuits.

\section{Formalism for the two-point impedance}\label{secII}
We first review the generic derivation of the expression for the two-point impedance in terms of the eigenvalues and eigenvectors of the circuit Laplacian~\cite{wu_theory_2004,tzeng_theory_2006,cserti_uniform_2011,izmailian_generalised_2014,izmailian_two-point_2014,cernanova_nonsymmetric_2014,mamode_revisiting_2021}. A $RLC$ circuit can be represented as a graph in which the vertices of the graph represent the voltage nodes and the edges represent the couplings between the nodes due to $R$, $L$, and $C$ components between the nodes. Under driving at a single AC frequency $\omega$, the circuit can be mathematically represented by its Laplacian matrix $J$, which relates the currents injected into the nodes with the voltages at each node via 
\begin{equation}
I = JV \label{LaplacianEq}
\end{equation}
where $I$ is a vector of the currents injected into each node and $V$ the corresponding vector of the node voltages. The Laplacian matrix for a circuit can be obtained simply by writing Kirchhoff’s current law at each voltage node. For example, consider a simple circuit consisting of two voltage nodes connected by a single capacitor with capacitance $C$. Applying Kirchhoff’s current law at the two nodes gives $I_1 = i\omega C (V_1-V_2)$ and $I_2 = i\omega C (V_2-V_1)$ where $I_a$ and $V_a$ are the injected current and voltage at node $a$, respectively. Using Eq.~\eqref{LaplacianEq} and these relations between the node current and voltages, the Laplacian matrix $J$ of this simple circuit is then given by $J = i\omega C \begin{pmatrix} 1 & -1 \\ -1 & 1 \end{pmatrix}.$ By definition, the impedance between two nodes $i$ and $j$ is the factor of proportionality between the voltage difference $V_i-V_j$ that develops between the two nodes when a current of magnitude $I$ is injected into node $i$ and extracted at node $j$: 
\begin{equation}
    Z_{ij}=\frac{V_i - V_j}{I}.
    \label{ImpZ}
\end{equation}
\noindent
To determine the impedance between two nodes, the voltages $V_i$ and $V_j$ have to be determined. To achieve this, we employ the circuit Green's function $G$, which is defined as the pseudo-inverse of the circuit Laplacian $G=J^{-1}$. Using the Green's function, Eq.~\eqref{LaplacianEq} can be rewritten as $V=GI$. The electrical voltage at node $i$ can thus be expressed as 
\begin{equation}
    V_i=\sum_{j}^{N} G_{ij} I_j,
    \label{Voltage_Greens}
\end{equation}
\noindent where $G_{ij}$ is the $(i,j)$th element of the pseudoinverse matrix $G$. By setting $I_i=I$ and $I_j=-I$ in Eq. \eqref{Voltage_Greens}, Eq.~\eqref{ImpZ} can be rewritten as 
\begin{equation}
    Z_{ij}=\sum_{k=i,j} \frac{G_{i k} I_k - G_{j k} I_k}{I},
    \label{Z_G}
\end{equation}
which gives
\begin{equation}
    Z_{ij}=G_{ii}+G_{jj}-G_{ij}-G_{ji}.
    \label{ZGreensFunc}
\end{equation}
\noindent 
By resolving in the space of eigenstates, $J$ can be written in terms of its right eigenvectors $|\psi_k\rangle$, left eigenvectors $\langle \psi_k |$, and eigenvalues $\lambda_k$ as $J = \sum_k |\psi_k\rangle \lambda_k \langle \psi_k|$. In general, $|\psi_k\rangle \neq \langle \psi_k|^\dagger$, because $J$ is not Hermitian. However, $|\psi_k\rangle = \langle \psi_k|^\dagger$ holds in the following special cases: (i) in a purely $LC$ circuit, where $J$ is anti-Hermitian and the $\lambda_k$'s are imaginary, and (ii) in a purely resistive circuit, where $J$ is Hermitian and the $\lambda_k$ are real.  For a general $RLC$ circuit, the $\lambda_k$ are complex, and the general relation $G=\sum_k |\psi_k\rangle (\lambda_k)^{-1} \langle \psi_k|$ holds. Note that because $G$ is defined as the \textit{pseudo}inverse of $J$, any zero eigenvalues are excluded from the sum if they exist.  
Writing $|\psi_k\rangle$ as a column vector of the node voltages $|\psi_k\rangle=\{\psi_{k 1},\psi_{k 2},\dots,\psi_{k N}\}$ where $N$ is the total number of nodes, Eq.~\eqref{ZGreensFunc} can rewritten as 
\begin{equation}
    Z_{ij}=\sum_{k, \lambda_k \neq 0} \frac{|\psi_{k i}-\psi_{k j}|^2}{\lambda_k},
    \label{Z_eig}
\end{equation}
\noindent where $\lambda_k$ is the corresponding eigenvalue of the Laplacian, and the $|...|$ norm is the biorthogonal norm. Any arbitrary two-point impedance can then be numerically calculated using Eq.~\eqref{ImpZ}, \eqref{ZGreensFunc}, or \eqref{Z_eig}.
\\
\\
\begin{figure*}[htp]
	\centering
	\includegraphics[width=\textwidth]{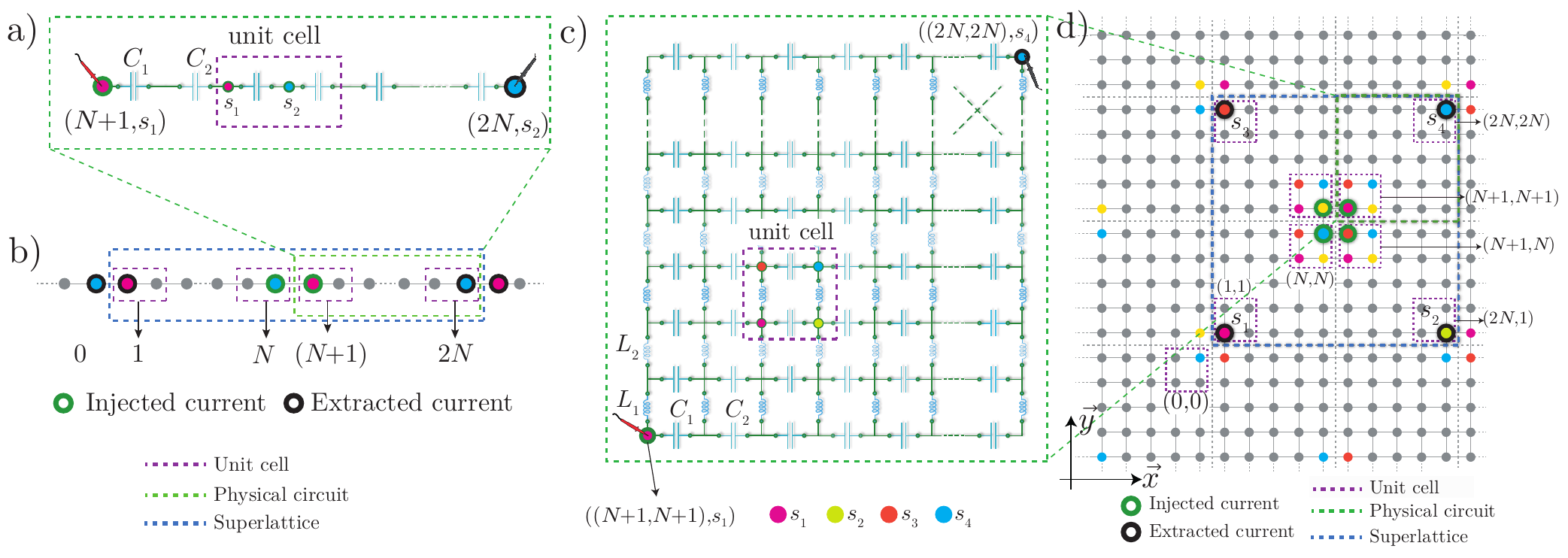}
	\caption{\textbf{Lattice structures of illustrative SSH circuits and their periodic image lattices for finite-lattice impedance computations}. (a) The 1D SSH circuit comprises two capacitors ($C_1$ and $C_2$) and two nodes ($s_1$ and $s_2$) in a unit cell. (b) The superlattice comprising the physical and image 1D SSH circuits is denoted by the blue dashed box, so constructed such that the impedance through a finite single block is recast into a problem with periodically placed injected and extracted currents. The green and purple dashed boxes denote the physical circuit and its unit cell, respectively. (c) The physical 2D SSH circuit consisting of two kinds of capacitors with capacitances $C_1$ (bold) and $C_2$ (thin) connecting the nodes along the horizontal direction and two kinds of inductors with inductances $L_1$ (bold) and $L_2$ (thin) linking the nodes along the vertical direction. The four distinct nodes in a unit cell (purple dashed square) $s_1$, $s_2$, $s_3$, and $s_4$ are represented as magenta, yellow, orange, and cyan circles, respectively. To measure the corner-to-corner impedance, current is injected at node $s_1$ of unit cell $(N+1,N+1)$ and extracted at node $s_4$ of unit cell $(2N, 2N)$ [also see the matrix representation in Eq.~\eqref{2DcurrentMatrix}]. (d) The infinite periodic lattice tiling of the $(2N)^2$-unit cell superlattice consisting the image and physical circuits (blue dashed square), which is constructed such that the impedance across the finite lattice can be expressed in terms of translation-invariant momentum contributions. The nodes at which current is injected and extracted are denoted with green and black outlines, respectively.}
	\label{topo_circuits} 
\end{figure*}

\section{Method of images for analytic impedance formulas across bounded circuit lattices}
In this section, we review and derive the general analytical formalism for calculating the impedance between two edge or corner nodes in finite discrete circuit lattices based on \textit{the method of images} and discuss their impedance behaviors. The method of images is needed to put finite lattice impedance computations, which are not easily represented by exact analytic formulas due to the lack of translation symmetry, on equal footing with periodic lattices. We shall illustrate the approach with exemplary circuit arrays with nontrivial unit cells, such that the tiling of the periodic images is not trivial.

In the context of electrostatic theory, the electric potential distribution inside an area of interest in the vicinity of a grounded conducting plate can be obtained by placing an image charge reflected across the conducting plate~\cite{riley_mathematical_1999,griffiths_introduction_2005,jackson_classical_1999,feynman1965feynman,yang_designing_2022,li_non-hermitian_2022,mamode_electrical_2017,rafi-ul-islam_interfacial_2022,pantoja_greens_2021}. The image charge causes the surface of the plate to become equipotential (which is considered zero for a grounded conductor) and thus satisfy the boundary conditions of a grounded conducting plate. 

Inspired by this approach, we apply an analogous idea by placing image circuits symmetrically about our desired circuit to replicate the boundary conditions satisfied by a finite circuit under open boundary conditions~\cite{mamode_electrical_2017,mead_resistance_2009}. %Unlike the classical electrostatic picture in which the tangential component of the electrostatic potential distribution is zero at the boundaries
The translation and inversion symmetries obeyed by the original and image circuits in our method force the voltage potentials at the boundary nodes of the original and the image circuits to be the same[see Fig.~\ref{VSpatial}(b)]. The equal potentials at the boundary nodes result in zero current flow across the boundaries between the physical and image circuits, and therefore replicates the boundary condition of zero current flow across the open physical boundaries of the finite bounded circuit. The potential distribution of the original circuit is then the same as that of the finite circuit with physical boundaries. This is the key insight that underlies our derivation of analytical expressions for the potential profiles of circuits with open boundaries via \textit{the method of images}. 

\subsection{Example: 1D SSH circuits}
To illustrate how \textit{the method of images} yields exact analytical expressions, we first consider sufficiently nontrivial circuit arrays in which each unit cell contains more than one type of node, such as the 1D and 2D topological Su-Schrieffer-Heeger (SSH) circuits. In such lattices, the transitions from the topologically trivial to non-trivial phases are accompanied by strong impedance resonances at the resonant frequency~\cite{lee_topolectrical_2018,helbig_band_2019,hofmann_chiral_2019,rafi-ul-islam_topoelectrical_2020,imhof_topolectrical-circuit_2018,2022arXiv220609931H,lenggenhager_simulating_2022,zhang2023electrical}. However, our focus here is on their edge-to-edge (1D SSH) and corner-to-corner (2D SSH) impedance profiles rather than the topological impedance characteristics.

The 1D SSH circuit consists of unit cells in which each unit cell contains two distinct nodes labeled as $s_1$ and $s_2$ where the nodes  are connected to each other by intra-cell capacitors of capacitance $C_1$ and inter-cell capacitors of capacitance $C_2$ [see Fig.~\ref{topo_circuits}(a)]. The corresponding $2 \times 2$ circuit Laplacian of a periodic 1D SSH circuit in momentum space is therefore written as
\begin{equation}
\mathcal{L}_{\text{1D}}^{\text{SSH}}(k_1) = i \omega \begin{pmatrix}
     C_1+C_2 & - C_1 - C_2 e^{-i k_1}\\ - C_1 - C_2 e^{i k_1} & C_1+C_2
\end{pmatrix},
\label{lap_1d_ssh}
\end{equation}
\noindent where $\omega$ is the driving AC frequency and $k_1$ is the crystal momentum along the $x$ direction. We label the nodes in the SSH chain as $\mathbf{\tilde{r}}=(\mathbf{r},\mu)$ where $\mu \in (s_1, s_2)$ denotes the node within each unit cell, $\mathbf{r}=n_1 \mathbf{a_1}$ denotes the location of the unit cell with $\mathbf{a_1}$ being the unit vector along the length of the chain and $n_1$ the coordinate of the unit cell, and the tilde on $\mathbf{\tilde{r}}$ denotes a composite index consisting of both the location of the unit cell and the sub-lattice site.

To investigate the behavior of the impedance as the circuit size increases, we consider the impedance between node $s_1$ in the leftmost unit cell and node $s_2$ in the rightmost unit cell in the 1D SSH chain circuit. Note that the circuit size is increased unit cell-by-unit cell so as to preserve lattice uniformity. Therefore, $N$ refers to the number of unit cells throughout this study.
Henceforth, we shall call the actual circuit with open boundaries (the green dashed rectangles in Fig.~\ref{topo_circuits}(b) and \ref{topo_circuits}(d)) \emph{the physical lattice} or \emph{the physical circuit}, the image(s) of the actual circuit \emph{the image lattice(s)} or \emph{the image circuit(s)}, and the block containing physical and image circuits (the blue dashed rectangles in Fig.~\ref{topo_circuits}(b) and \ref{topo_circuits}(d)) \emph{the superlattice} or \emph{the supercircuit}. To obtain the voltages in a finite SSH chain containing $N$ unit cells that result from current injection at the left-most node and current extraction at the right-most node, we place an image chain of the same length to the left of physical chain circuit, but importantly reflected about the chain boundary so that no current flows across it due to reflection symmetry. We then inject current at the right-most node of the image circuit and the left-most node of the physical circuit,  and extract the injected current at the left-most node of the image circuit and the right-most node of the physical circuit, as shown in Fig.~\ref{topo_circuits}(b).  The impedance between any two lattice points can then be obtained using by Eq.~\eqref{ImpZ} once the voltage distribution is known. 

To determine the voltage distribution explicitly, we utilize Ohm's law, which states that the current distribution $\mathbfcal{J} = \sigma \mathbf{E}$ where the electric field is given by $\mathbf{E}=-\grad V$. Therefore, the current distribution can be written as $\mathbfcal{J}=-z^{-1} \grad V$, where $z$, the uniform impedance between each node, is the inverse of the electrical conductivity $\sigma$. Owing to Kirchhoff's current law, the current density is also written as $\div{\mathbfcal{J}}= I \left( \delta_{(\mathbf{r'},\nu)\in\mathbf{\tilde{r}_{in}}}(\mathbf{r},\mu,\mathbf{r'},\nu)-\delta_{(\mathbf{r'},\nu)\in\mathbf{\tilde{r}}_{out}}(\mathbf{r},\mu,\mathbf{r'},\nu)\right)$ where $\delta$ denotes the Kronecker delta and $I$ represents the current magnitude. After performing the relevant substitutions, we arrive at a Poisson-type equation. Here, because the Green's function satisfies $\grad^2 G(\mathbf{r,\mu,r',\nu}) = -\delta(\mathbf{r,\mu,r',\nu})$ by definition \cite{katsura_lattice_1971,cserti_application_2000,guttmann_lattice_2010,joyce_exact_2002,joyce1973simple,montroll_random_1965} (note that we have rewritten the circuit Green's function $G$ of Sect.~\ref{secII} in the quasi-continuum picture to facilitate the discussion), the voltage at sub-lattice node $\mu$ of the unit cell at location $\mathbf{r}$ is found as
\begin{equation}
    \begin{aligned}
V(\mathbf{r},\mu) = I \Big( & \sum_{(\mathbf{r'},\nu
) \in \mathbf{\tilde{r}}_{in}} G(\mathbf{r},\mu,\mathbf{r'},\nu) \\
    & - \sum_{(\mathbf{r'},\nu)\in \mathbf{\tilde{r}}_{out}} G(\mathbf{r},\mu,\mathbf{r'},\nu)  \Big),
    \end{aligned}
    \label{volt_distr}
\end{equation}
\noindent where 
\begin{equation}
\begin{aligned}
	\mathbf{\tilde{r}}_{\text{in}} &\in \{(N\mathbf{a}_1,s_2),((N+1) \mathbf{a}_1,s_1)\},  \\
\mathbf{\tilde{r}}_{\text{out}} &\in \{(1\mathbf{a}_1 ,s_1),(2N \mathbf{a}_1,s_2)\} 
  \label{current_1dssh}
\end{aligned}
\end{equation}
denote the nodes where the currents are injected ($\mathbf{\tilde{r}}_{in}$) and extracted ($\mathbf{\tilde{r}}_{out}$). The spatial Green's function $G(\mathbf{r},\mu,\mathbf{r'},\nu)$ in Eq.~\eqref{volt_distr} can be determined by applying the discrete Fourier transform given by
\begin{equation}
	G(\mathbf{r},\mu,\mathbf{r'},\nu) = \frac{1}{(2N)^D} \sum_\mathbf{k} G(\mathbf{k})_{[\mu,\nu]} e^{i \mathbf{k}\cdot(\mathbf{r} - \mathbf{r'})},
	\label{GreenFourier}
\end{equation}
\noindent where the momentum space index $\mathbf{k}=k_1 \mathbf{a_1}$ in which $k_1 = n_1 \pi/N$ and $n_1$ varies over a $2N$ period and $G(\mathbf{k})_{[\mu,\nu]}$ is the matrix element of the pseudoinverse of the momentum-space circuit Laplacian [for this example, it is $\mathcal{L}_{\text{1D}}^{\text{SSH}}(k_1)$ given in Eq.~\eqref{lap_1d_ssh}]. $D$ represents the circuit dimension. We then tile the $2N$ unit cells comprising the physical and image circuits to form an infinite-sized lattice with a period of $2N$. Therefore, the Green’s function and the circuit Laplacian are constructed for the superlattice with a period of $2N$ such that the symmetric current injections and extractions in this periodic infinite lattice lead to a symmetric spatial voltage distribution with the period of $2N$; hence, the current entering the physical circuit cannot leak out through the boundaries of the physical circuit. Accordingly, the open boundary condition for the physical circuit with $N$ unit cells is satisfied. To realize this, we use the current distribution Eq.~\eqref{current_1dssh} with Eq.~\eqref{volt_distr} to determine the voltage at the leftmost node of the physical circuit as
\begin{equation}
    \begin{aligned}
    	 V(\mathbf{r}=N+1,\mu=s_1) =  I \big(& G((N+1),s_1, (N),s_2)\\& + G((N+1),s_1, (N+1),s_1)\\
& - G((N+1),s_1, (1),s_1) \\& - G((N+1),s_1,(2N),s_2) \big),
    \end{aligned}
    \label{1Dssh_Vin}
\end{equation}
\noindent and that at the rightmost edge node as
\begin{equation}
    \begin{aligned}
    	V(\mathbf{r}=2N,\mu=s_2)  = I \big(& G((2N),s_2, (N),s_2) \\&+ G((2N),s_2, (N+1),s_1) \\
				&-G((2N),s_2, (1),s_1) \\ &-G((2N),s_2,(2N),s_2) \big).
    \end{aligned}
    \label{1Dssh_Vout}
\end{equation}
To find the voltages explicitly, we insert the momentum-space Green's function given in Eq.~\eqref{GreenFourier} into Eqs.~\eqref{1Dssh_Vin} and ~\eqref{1Dssh_Vout} so that the impedance between the two edge nodes can be calculated as $Z_{1D}^{SSH}=V((N+1),s_1)-V((2N),s_2)$. At this point, we utilize the inversion and translation symmetries (i.e., $J(\mathbf{r})=J(\mathbf{-r})$ and $J(\mathbf{r})=J^{\intercal}(\mathbf{r})$ (where $(^\intercal)$ denotes the transpose operation), respectively) that our circuit possesses~\cite{venezian_resistance_1994,owaidat_resistance_2013,perrier_symmetries_2021,joyce1973simple,montroll_random_1965}. Because of the infinite periodic tiling, these symmetries imply that the voltage at the nodes where current is injected has the same magnitude as that at the nodes where current is extracted but with the opposite sign, i.e., $V(\mathbf{\tilde{r}} \in \mathbf{\tilde{r}}_{in}) = -V(\mathbf{\tilde{r}} \in \mathbf{\tilde{r}}_{out})$. Therefore, by using these symmetries and performing the relevant substitutions, the voltages at the edge nodes are found to be
\begin{equation}
\begin{aligned}
		V(&(N+1),s_1) = - V((2N),s_2) = \frac{I}{4N} \sum_{n_1=1}^{2N} \times  \\
		  & \frac{(e^{i n_1 \pi}-1)\left(C_2(1-e^{-i n_1 \pi})+C_1\left( 1-e^{-i n_1 \pi} e^{i n_1 \pi/N}\right) \right)}{i\omega C_1 C_2 \left(1-\cos (n_1 \pi /N)\right)},
\end{aligned}
\label{volt1DSSH}
\end{equation}
where we introduced $\mathbf{r}=n_1\mathbf{a_1}$ and $\mathbf{k}=k_1\mathbf{a_1}$ where $k_1 = n_1 \pi/N$. Here, because $e^{i n_1 \pi}-1=0$ when the integer $n_1$ is $even$, the summation is performed only for $odd$ $n_1$s. By considering $Z_{1D}^{SSH}=2V((N+1),s_1)/I=-2V((2N),s_2)/I$, and by means of trigonometric conversions [e.g., $1+e^{i n_1 \pi /N}=1+\cos (n_1\pi/N)+i \sin (n_1\pi/N)$ where the sine function can be neglected because of its zero contribution to the real part of the impedance], the edge-to-edge impedance as a function of the circuit size $N$ is given by
\begin{equation}
	Z_{1D}^{\text{SSH}}(N) = \frac{1}{N} \sideset{}{^*} \sum_{k_1} \frac{2C_2/C_1+(1+\cos k_1 )}{-i \omega C_2(1-\cos k_1 )}.
	\label{Imp1dsshCap}
\end{equation}
\noindent As before, $k_1 = n_1 \pi/N$ where $n_1$ is varied over the superlattice, i.e., $n_1\in\{1,2,\dots,2N\}$; however, the summation is restricted over $n_1 \in odd$ because of the above-mentioned summation rule, which the asterisk ($*$) on the summation operator indicates.
\\
\indent Using the same procedure, the edge-to-edge impedance in a circuit where the $C_1$ capacitors are replaced by capacitors with capacitance $C$ and the $C_2$ capacitors by inductors with inductance $L$ is obtained as
\begin{equation}
    Z_{1\text{D}}^{\text{SSH}_L}(N) = \frac{1}{N} \sideset{}{^*} \sum_{k_1} \frac{2- \omega^2 C L(1+\cos k_1)}{-i \omega C(1-\cos k_1)}.
	\label{Imp1dsshInd}
\end{equation}
%Note that, the same notation such as $k_1$ and the summation regulation in Eq.~\eqref{Imp1dsshCap} apply to the above equation.
These formulas provide the impedance between nodes $s_1$ and $s_2$ in the leftmost and rightmost unit cells, respectively. Note that our sum includes a total of $N$ points, which corresponds to the $N$ unit cells.
\newline
\\
\subsection{Example: 2D SSH circuit}
We now proceed with a higher-dimensional circuit in which all the principal directions are non-trivial. For example, a 2D SSH circuit can be constructed by extending the 1D SSH circuit along the new $y$ direction. We first consider a capacitive 1D SSH circuit with intracell coupling $C_1$ and intercell coupling $C_2$. We then extend the circuit along the $y$ direction by using two inductances $L_1$ and $L_2$ to connect the nodes along the vertical direction, as shown in Fig.~\ref{topo_circuits}(c). The resultant unit cell [dashed purple square in Fig.~\ref{topo_circuits}(c)] has four distinct nodes denoted as $s_1$ to $s_4$ in which nodes $s_1$ and $s_4$ are located at opposite corners of the unit cell. Since preserving the uniformity of the unit cells requires the circuit size to be increased in multiples of the unit cells, the corner-to-corner impedance is measured between node $s_1$ in the first unit cell and node $s_4$ in the unit cell at the opposite corner of the 2D SSH circuit. The circuit Laplacian for a periodic 2D SSH circuit in momentum space is written as
\begin{equation}
    \mathcal{L}_{2\text{D}}^{\text{SSH}}(k_1,k_2) = i\omega \begin{pmatrix}
    \Sigma & \Gamma & \Delta & 0\\
    \Gamma^* & \Sigma & 0 & \Delta \\
    \Delta^* & 0 & \Sigma & \Gamma\\
    0 & \Delta^* & \Gamma^* & \Sigma
\end{pmatrix},
\label{2d_ssh_lap}
\end{equation}
\noindent where $\Gamma= -C_1 - C_2 e^{- i k_1} $, $\Delta= \frac{1}{\omega^2 L_1} + \frac{1}{\omega^2 L_2} e^{-i k_2}$, and $\Sigma = C_1 + C_2 - \frac{1}{\omega^2 L_1} - \frac{1}{\omega^2 L_2}$. Similar to the 1D SSH circuit, we evaluate the voltage distribution for the impedance measurement by introducing image circuits around the physical circuit [Fig.~\ref{topo_circuits}(d)], injecting current at $\mathbf{\tilde{r}}_{\text{in}}$, and extracting current at $\mathbf{\tilde{r}}_{\text{out}}$ where 
\begin{widetext}
\begin{equation}
\begin{aligned}
   \mathbf{\tilde{r}}_{\text{in}} &\in \{ ((N+1,N+1),s_1),((N,N+1),s_2),((N+1,N),s_3),((N,N),s_4)\}  \\
    \mathbf{\tilde{r}}_{\text{out}} &\in \{ ((1,1),s_1),((2N,1),s_2),((1,2N),s_3),((2N,2N),s_4)\}.
  \label{2dssh_current}
\end{aligned}
\end{equation}
\end{widetext}
\noindent Here, the spatial positions of the nodes at which current is injected and extracted are written in the form of $((n_1\mathbf{a_1},n_2 \mathbf{a_2}),s_\alpha)$ where $\alpha=(1,2,3,4)$ and $n_2$ and $\mathbf{a_2}$ are the coordinate and unit vector along the $y$ direction, respectively. 
The voltage $V\big(\mathbf{r}=(N+1,N+1),\mu=s_1\big)$ at the lower left corner of the physical circuit [see Fig.~\ref{topo_circuits}(d)] is thus given by \\
\\
\begin{widetext}
	\begin{equation}
		\begin{aligned}
			 V\Big(&\mathbf{r}=(N+1,N+1),\mu=s_1\Big) =\\
			 I \Big(&G\big((N+1,N+1),s_1,(N+1,N+1),s_1\big) + 
			 G\big((N+1,N+1),s_1,(N,N+1),s_2\big) \\
			 &+G\big((N+1,N+1),s_1,(N+1,N),s_3\big) + 
			 G\big((N+1,N+1),s_1,(N,N),s_4\big) -
			 G\big((N+1,N+1),s_1,(1,1),s_1\big) \\
			&-G\big((N+1,N+1),s_1,(2N,1),s_2\big) - 
			 G\big((N+1,N+1),s_1,(1,2N),s_3\big) - 
			 G\big((N+1,N+1),s_1,(2N,2N),s_4\big) \Big).
		\end{aligned}
		\label{2dssh_V1}
	\end{equation}
\noindent Similarly, the voltage at the upper right corner of the physical circuit is given by
	\begin{equation}
		\begin{aligned}
			 V\Big(&\mathbf{r}=(2N,2N),\mu=s_4\Big) = \\
			 I \Big(& G\big((2N,2N),s_4,(N+1,N+1),s_1\big) + 
			 G\big((2N,2N),s_4,(N,N+1),s_2\big) + 
			 G\big((2N,2N),s_4,(N+1,N),s_3\big) \\
			&+G\big((2N,2N),s_4,(N,N),s_4\big) -
			 G\big((2N,2N),s_4,(1,1),s_1\big) - 
			 G\big((2N,2N),s_4,(2N,1),s_2\big) \\
			&-G\big((2N,2N),s_4,(1,2N),s_3\big) -
			 G\big((2N,2N),s_4,(2N,2N),s_4\big) \Big).
		\end{aligned}
		\label{2dssh_V2}
	\end{equation}
\end{widetext}

\noindent We now employ the discrete Fourier transform given in Eq.~\eqref{GreenFourier} to evaluate Eqs.~\eqref{2dssh_V1} and \eqref{2dssh_V2} explicitly. Because of the aforementioned circuit symmetries, the potential difference between the nodes at two opposite corners when the nodes are connected by a current source is $V((N+1,N+1),s_1) - V((2N,2N),s_4) = 2 V((N+1,N+1),s_1) = -2 V((2N,2N),s_4)$. By substituting Eq.~\eqref{GreenFourier} into Eq.~\eqref{2dssh_V1} [Eq.~\eqref{2dssh_V2}], the voltage at the lower left (upper right) corner node can be obtained. The two-point impedance between opposite corner nodes is given by $Z_{2D}^{\text{SSH}}(N) = 2 V((N+1,N+1),s_1)$ or $Z_{2D}^{\text{SSH}}(N) = -2 V((2N,2N),s_4)$, for which we provide the full impedance expression in Appendix \ref{appendix_2dssh_formula}. Although the analytical expression looks complicated, it demonstrates the utility of the method of images technique. The resulting voltage distribution over the superlattice reflects the inversion and translation symmetries of the circuit. We now present an example of the spatial voltage distribution in the superlattice containing the physical 2D SSH circuit and its image copies.

\subsubsection{Spatial voltage distribution of the 2D SSH circuit}

To show how the symmetric current injection and extraction over a periodic superlattice gives rise to equal potentials between the boundary nodes of the image and physical circuits, we present the spatial voltage distribution for a periodic 2D SSH circuit. The spatial voltage distribution can be calculated using the numerical Laplacian formalism [Eq.~\eqref{LaplacianEq}]
\begin{equation}
	V=(J_{2\text{D}\text{-SSH}}^{\text{periodic}})^{-1} I_{\text{2D-SSH}},
\end{equation}
where the voltage matrix ($V$) over the periodic superlattice is obtained by performing the matrix multiplication of the inverse Laplacian matrix ($J_{2D\text{-SSH}}^{\text{periodic}}$) and the current matrix ($I_{\text{2D-SSH}}$). The current matrix corresponding to the 2D SSH circuit shown in Fig.~\ref{topo_circuits}(d) is written as
\begin{equation}
	I_{2D\text{-SSH}}=I\left(\begin{NiceArray}{>{\strut}cccccccccc}[margin,extra-margin = 0pt]
		-1 & 0 & \cdots & 0 & 0 & 0 & 0 & \cdots & 0 & -1\\
		0 & 0 & \cdots & 0 & 0 & 0 & 0 & \cdots & 0 & 0\\
		\vdots & \vdots & \ddots & \vdots & \vdots  & \vdots & \vdots & \ddots & \vdots & \vdots  \\
		0 & 0 & \cdots & 0 & 0 & 0 & 0 & \cdots & 0 &0\\
		0 & 0 & \cdots & 0 & 1 & 1 & 0 & \cdots & 0 &0\\
		0 & 0 & \cdots & 0 & 1 & 1 & 0 & \cdots & 0 &0\\
		0 & 0 & \cdots & 0 & 0 & 0 & 0 & \cdots & 0 &0\\
		\vdots & \vdots & \ddots & \vdots & \vdots  & \vdots & \vdots & \ddots & \vdots & \vdots  \\
		0 & 0 & \cdots & 0 & 0 & 0 & 0 & \cdots & 0 &0\\
		-1 & 0 & \cdots & 0 & 0 & 0 & 0 & \cdots & 0 & -1\\
		\CodeAfter
		\begin{tikzpicture}
			\node [draw=red, rounded corners=3pt, inner ysep = 2pt,
			fit = (5-6) (1-10)] {} ;
		\end{tikzpicture}
	\end{NiceArray}\right)_{2N\times 2N}
	\label{2DcurrentMatrix}
\end{equation}

\begin{figure}[h!]
	\centering
	\includegraphics[width=8.5cm]{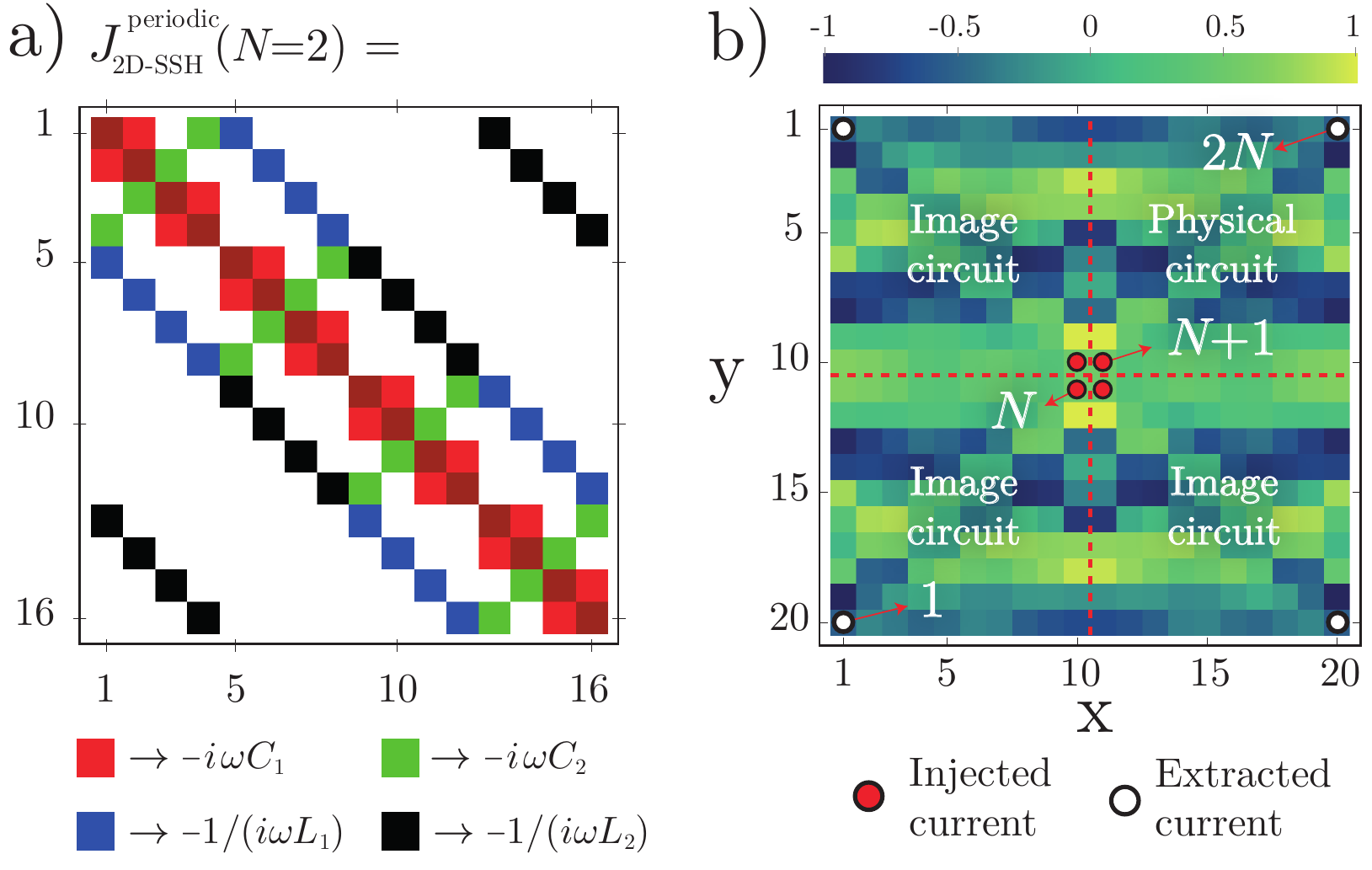}
	\caption{\textbf{The matrix representation of the periodic circuit Laplacian of the 2D SSH circuit for $N=2$ and its spatial voltage distribution for $N=10$.} (a) While the admittances between the node links are represented by the red, green, blue, and black squares in the off-diagonal elements, the darker red squares in the diagonal elements represent the total node conductance. %such that $\raisebox{.5ex}{\fcolorbox{white}{red}{\rule{0pt}{2pt}\rule{2pt}{0pt}}} \mathbf{\rightarrow} -i\omega C_1$, $\raisebox{.5ex}{\fcolorbox{white}{green}{\rule{0pt}{2pt}\rule{2pt}{0pt}}} \mathbf{\rightarrow} -i\omega C_2$, $\raisebox{.5ex}{\fcolorbox{white}{blue}{\rule{0pt}{2pt}\rule{2pt}{0pt}}} \mathbf{\rightarrow} -1/(i\omega L_1)$, $\raisebox{.5ex}{\fcolorbox{white}{black}{\rule{0pt}{2pt}\rule{2pt}{0pt}}} \mathbf{\rightarrow} -1/(i\omega L_2)$ 
	(b) Spatial voltage distribution matrix of the 2D SSH circuit for $N=10$ presented as a density plot. The red dashed lines separating the entire matrix into four blocks represent the boundaries between the physical and image circuits. The colors of the squares represent the magnitude of the voltage potential. Identical colors on both sides of the red dashed lines imply that there are zero potential differences between the boundary nodes. The red and white circles with black frames represent the nodes where the current is injected and extracted, respectively.}
	\label{VSpatial}
\end{figure}

The matrix elements framed by the red square in the upper right block corresponds to the current matrix of the physical circuit. Because we consider an infinite lattice tiling with a superlattice with a total period of $2N$ unit cells along each direction, we employ the periodic circuit Laplacian ($J_{2D\text{-SSH}}^{\text{periodic}}$). In Fig.~\ref{VSpatial}(a), we display an example of the periodic circuit Laplacian of the 2D SSH circuit for $N=2$. Therefore, the matrix multiplication of the periodic Laplacian and the current matrix yields the spatial voltage distribution of the 2D SSH circuit. An example for the voltage distribution when $N=10$ is given in Fig.~\ref{VSpatial}(b) where the voltage matrix is presented as a density plot. As can be seen from Fig.~\ref{VSpatial}(b), the boundary nodes of the physical circuit have the same voltages as those of the boundary nodes of the image circuits. Due to the equal voltage potentials between the boundary nodes, the current injected at node $N+1$ cannot flow into the image circuits and is instead contained within the physical circuit. Therefore, the symmetrical current engineering as an analogy of the method of images leads to a perfectly symmetrical voltage distribution that therefore satisfies the boundary conditions. This makes it possible to obtain analytical expressions for finite-size circuits by applying the method of images to an infinite periodic lattice.

\section{Simplified analytic impedance formulas for RLC circuits with a single node per unit cell}\label{secRLC}

\begin{figure*}[t]
	\centering
	\includegraphics[width=\textwidth]{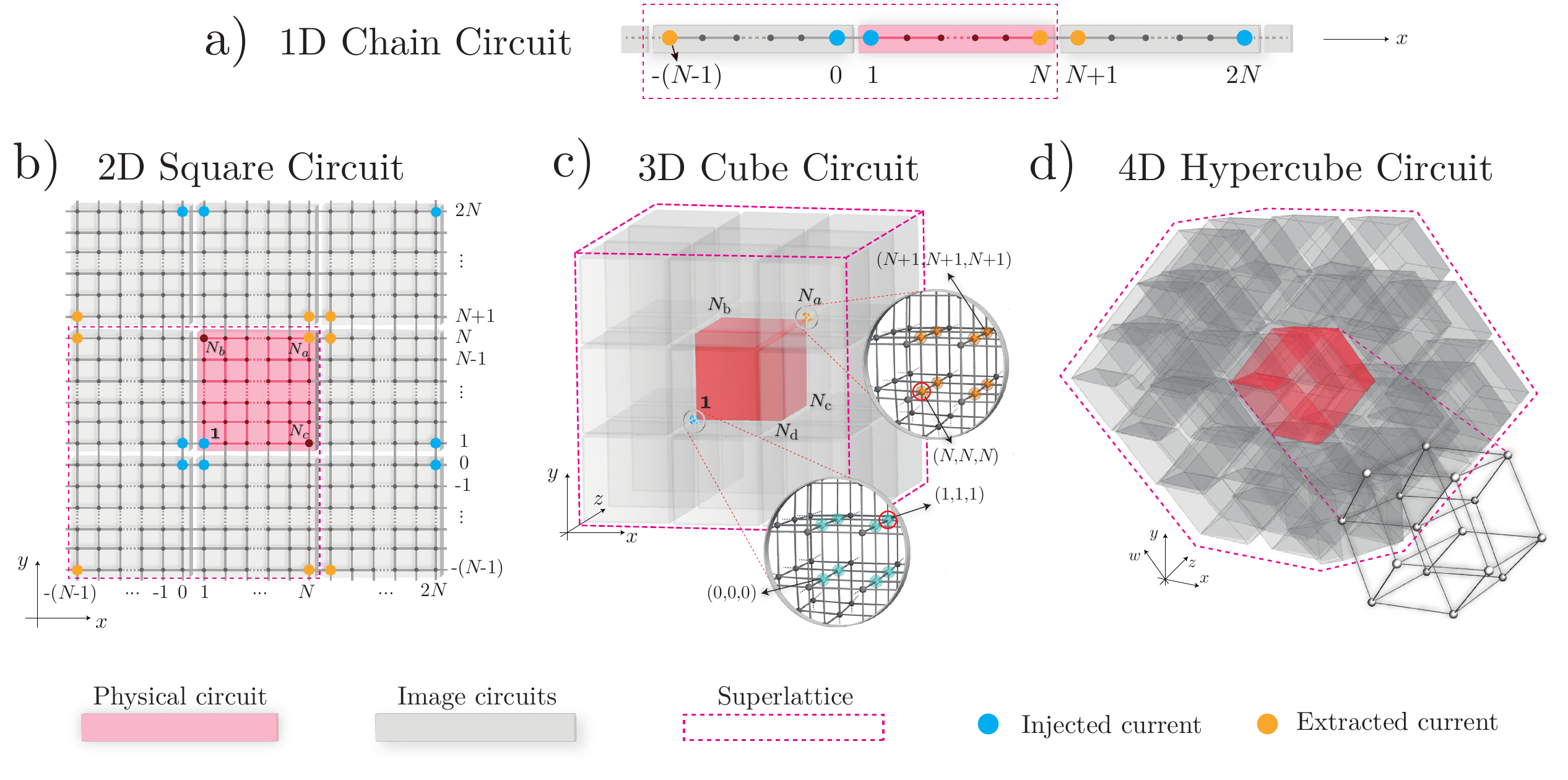}
	\caption{\textbf{Circuits lattices of different dimensionalities and their method of images implementations.} The impedance is measured (a) edge-to-edge in the 1D circuit and (b)-(d) corner to corner in the 2D and higher-dimensional circuits. Red and gray blocks represent the physical and image circuits constructing a periodic infinite lattice. Each superlattice consisting of a physical circuit and image circuits is marked by the magenta dashed lines in each illustration. The cyan and orange dots indicate the spatial positions of the nodes at which the current is injected and extracted, respectively. We label the corner nodes in the 2D square and 3D cube circuits as $\mathbf{N}_\alpha$ such that $\mathbf{N}_a$ is the diagonally opposite corner node to node $\mathbf{1}$ in every circuit. (b) The bigger red dots on the red plate denote the corner nodes $\mathbf{N}_c$ and $\mathbf{N}_b$ along the $x$ and $y$ directions, respectively. (c) The lower (upper) gray circle focuses on the node clusters where the current is injected (extracted). The red circles in the gray circles indicate the nodes belonging to the physical circuit i.e., nodes $\mathbf{1}$ and $\mathbf{N_a}$. (d) An illustrative representation of the 4D hypercube circuit. The input and output currents are not drawn to avoid excessive clutter. These circuits are homogeneous when the links are all resistive with admittances of $1/R_i$, all capacitive with admittances of $ z_{i} = i\omega C_{i} $, or all inductive with admittances $z_{i}= 1/(i\omega L_{i})$ [here $i=(1,2,\dots,D)$], but are heterogeneous when at least two distinct admittances with opposite phases such as $z_{1}=i\omega C$ and $z_{2}= 1/(i\omega L)$ are present.}
	\label{network_schematics}
\end{figure*}
 
Here, we derive a generalized analytical expression for the impedance between two nodes at the opposite edges in 1D and at opposite corners in higher dimensions in homogeneous and heterogeneous finite circuits that have a single node per unit cell. The nodes are connected by components with an admittance of $z_\alpha$ along the $\alpha$th direction where $\alpha \in  (1,2,\cdots, D)$ where $D$ is the dimension of the circuit. We define a homogeneous circuit as one in which the $z_\alpha$'s along all the directions have the same phase, i.e., they are either all capacitors or all inductors, and a heterogeneous circuit as one in which the $z_\alpha$'s along the different directions have different phases. To derive a generic expression, let us first consider a 2D infinite lattice made of image copies of a physical finite circuit [refer to Fig.~\ref{network_schematics}(b)] where the current is injected at nodes $\mathbf{r}_{\text{in}}$ and extracted from nodes $\mathbf{r}_{\text{out}}$. Using the definition of the Green's function $G J = -\delta $ [Eq.~\eqref{LaplacianEq}] and the translation invariance of the circuit [which implies that $G(\mathbf{r},\mathbf{r'}) = G (\mathbf{r}-\mathbf{r'})$], the voltage at any lattice point $\mathbf{r}$ is found to be
\begin{equation}
	V(\mathbf{r}) = I \left( \sum_{\mathbf{r'} \in \mathbf{r}_{\text{in}}} G(\mathbf{r - r'}) - \sum_{\mathbf{r'} \in \mathbf{r}_{\text{out}}} G(\mathbf{r - r'}) \right),
	\label{voltage_distribution}
\end{equation}
\noindent where $\mathbf{r} = n_1 \mathbf{a_1}+ n_2 \mathbf{a_2}$ with integers $(n_1,n_2) \in \{-(N-1),\cdots,N\}$, and $\mathbf{a_1}$ and $\mathbf{a_2}$ are unit vectors corresponding to the $x$ and $y$ directions, respectively. Note that we define the limits of the superlattice as $\{-(N-1),\cdots,N\}$ by choice unlike the limits of the superlattice that we define for the 1D and 2D SSH circuits. The period of the superlattice remains unchanged at $2N$. We now perform the discrete Fourier transformation [i.e., $G(\mathbf{r-r'})=1/(2N)^D \sum_{\mathbf{k}} G(\mathbf{k}) e^{i \mathbf{k}.(\mathbf{r-r'})}$] and recall $G(\mathbf{k})=\mathcal{L}^{-1}(\mathbf{k})$ to determine the spatial Green's function $G(\mathbf{r-r'})$ in Eq.~\eqref{voltage_distribution} as
\begin{equation}
	G(\mathbf{r-r'})=\frac{1}{(2N)^2}\sum_{\mathbf{k}}{\frac{e^{i \mathbf{k \cdot (r-r')}}}{\mathcal{L(\mathbf{k})}}},
	\label{Green_r}
\end{equation}
\noindent where $\mathbf{k}=(k_1 \mathbf{a_1} + k_2 \mathbf{a_2})$ is the momentum space index where $k_i=n_i\pi/N$ where $i=(1,2)$ and $n_i\in\{-(N-1),\dots,N\}$. Notice that, because we derive an analytical expression for the circuit that has a trivial unit cell [i.e., the circuit is made of a single-type node], the momentum-space Laplacian can take the role of momentum-space Green's function since $G(\mathbf{k})$ is no longer a matrix but is instead just the reciprocal of $\mathcal{L}(\mathbf{k})$. To achieve a symmetric voltage distribution over the superlattice, the current is injected and extracted at the nodes as depicted in Fig.~\ref{network_schematics}(b):
\begin{equation}
	\begin{aligned}
		&\mathbf{r}_{in} \in \{(0,0),(1,0),(0,1),(1,1)\}  \\
		&\mathbf{r}_{out} \in \{(N,N),(N+1,N),(N,N+1),(N+1,N+1)\},
	\end{aligned}
	\label{current_cases}
\end{equation}
where $(n_1,n_2)$ is a short-hand notation for $\mathbf{r}=(n_1\mathbf{a_1}+n_2\mathbf{a_1})$. From Eqs.~\eqref{current_cases} and \eqref{voltage_distribution}, the node voltage V($\mathbf{r}$) can be found through
\begin{equation}
	\begin{aligned}
		V(\mathbf{r}) = I  \Big( & G(\mathbf{r}) + G(\mathbf{r}+\mathbf{a_1}) + G(\mathbf{r}+\mathbf{a_2}) + G(\mathbf{r}+\mathbf{u}) \\
		& -G(\mathbf{r}+N\mathbf{u}) - G(\mathbf{r}+N\mathbf{u}+\mathbf{a_1}) \\ & -G(\mathbf{r}+N\mathbf{u}+\mathbf{a_2}) -
		G(\mathbf{r}+N\mathbf{u}+\mathbf{u}) \Big),
	\end{aligned}
	\label{voltage_final_with_green}
\end{equation}
where $\mathbf{u} = \mathbf{a_1}+\mathbf{a_2}$ denotes the unit vector for 2D lattices. To proceed, we insert the Green's function defined in Eq.~\eqref{Green_r} into Eq.~\eqref{voltage_final_with_green} to obtain the voltages at crosswise nodes $\mathbf{1}$ and $\mathbf{N}_a$ where $\mathbf{1}=1\mathbf{a_1}+1\mathbf{a_2}$ and $\mathbf{N}_a = N\mathbf{a_1}+N\mathbf{a_2}$. As mentioned, the translation symmetry in our circuit implies that $V(\mathbf{r})=-V(\mathbf{-r})$. Thus, the voltage difference between nodes $\mathbf{1}$ and $\mathbf{N}_a$ is $V(\mathbf{1})-V(\mathbf{N}_a) = 2 V(\mathbf{1}) = -2 V(\mathbf{N}_a)$. Therefore, we find the voltage at the corner nodes as
\begin{equation}
	\begin{aligned}
	&V(\mathbf{1} ) = - V(\mathbf{N}_\alpha)=  \frac{I}{4 N^2}\sum_{n_1} \sum_{n_2} \frac{1}{\mathcal{L}(\mathbf{k})} \times \\ & \left(1-e^{i\pi(n_1+n_2)}\right) \left(1+e^{i \pi n_1/N}+e^{i \pi n_2/N}+e^{i \pi (n_1+n_2)/N}\right).
		\end{aligned}
	\label{volt_final_exp}
\end{equation}
Notice that the term $(1-e^{i\pi (n_1+n_2)})$ in the numerator of Eq.~\eqref{volt_final_exp} is zero when $(n_1+n_2)$ is $even$ and is 2 when $(n_1+n_2)$ is $odd$. Therefore, this term can be replaced by 2 provided that the summation is restricted to odd $(n_1+ n_2)$. After simplifying Eq.~\eqref{volt_final_exp}, the impedance between the corner nodes $\mathbf{1}$ and $\mathbf{N}_a$ in a 2D square circuit, as a function of circuit size $N$, can be written as
\begin{equation}
	\begin{aligned}
		Z_{2\text{D}}^{\mathbf{1,N_a}}(N)&= \frac{2}{N^2} \sideset{}{^*} \sum_{\mathbf{k}} \\ 
& \times	\frac{\cos(k_{1}/2)\cos(k_2/2)\cos((k_1+k_2)/2)}{\mathcal{L}(\mathbf{k})},
	\end{aligned}
	\label{final_imp}
\end{equation}
where the asterisk on the sum operator indicates that $\mathbf{k}=\sum_{i} k_i \mathbf{a}_i$ where $k_i=n_i \pi /N$ and $i=(1,2)$ and a restricted summation over odd $(n_1+n_2)$. $\mathcal{L}(\mathbf{k})$ represents the corresponding 2D circuit Laplacian. One can then calculate the two-point impedance for both 2D homogeneous and heterogeneous circuits by simply assigning the corresponding circuit Laplacian to Eq.~\eqref{final_imp}. This equation is also valid for the impedance between any pair of corner nodes as long as the summation is taken over $n_2 \in \text{odd}$ for $Z_{2\text{D}}^{\mathbf{1,N_b}}$ and $n_1 \in \text{odd}$ for $Z_{2\text{D}}^{\mathbf{1,N_c}}$. This is because in the derivation of the expressions for $Z_{2\text{D}}^{\mathbf{1,N_b}}$ or $Z_{2\text{D}}^{\mathbf{1,N_c}}$, one arrives at Eq.~\eqref{volt_final_exp} with the factor $(1-e^{i n_2\pi})$ in the numerator for  $Z_{2\text{D}}^{\mathbf{1,N_b}}$ and $(1-e^{i n_1\pi})$ for $Z_{2\text{D}}^{\mathbf{1,N_c}}$ because, for example, the current distribution when the current is injected and extracted at nodes $\mathbf{1}$ and $\mathbf{N_c}$ is written as $\mathbf{r}_{\text{in}} \in \{(0,0),(1,0),(0,1),(1,1)\}$ and $\mathbf{r}_{\text{out}} \in \{(N,0),(N+1,0),(N,1),(N+1,1)\}$, respectively. (Here, $\mathbf{N}_b = N\mathbf{a_1}$ and $\mathbf{N}_c =N\mathbf{a_2}$ are the vertical and horizontal opposite corner nodes to the lower left node $\mathbf{1} =1\mathbf{a_1}+1\mathbf{a_2}$, respectively [see Fig.~\ref{network_schematics}(b)].) Therefore, there are contributions to the impedance only when $n_1\in \text{odd}$ for $Z_{2\text{D}}^{\mathbf{1,N_c}}$ or $n_2\in \text{odd}$ for $Z_{2\text{D}}^{\mathbf{1,N_b}}$. From here, we can deduce that the Fourier component(s) of the principal direction(s) corresponding to the corner node only contribute to the impedance when its (their) summation is odd.

Inspired by the derivation for the impedance formula for the 2D square circuit (Eq.~\eqref{final_imp}) and taking into account the summation rule for the impedance between different corner nodes, we can obtain a general analytical expression for the corner-to-corner impedance of both $D$-dimensional homogeneous and heterogeneous circuits as
\begin{equation}
	Z(N)=\frac{2}{N^D} \sideset{}{^*} \sum_{\mathbf{k}} \frac{\left( \prod_{i=1}^D \cos(k_i /2)\right) \times \cos \left(\sum_{i=1}^D k_i /2 \right) }{\left(\sum_{i=1}^D \lambda_{i}(1-\cos(k_i)\right)+z_{gnd}/2},
	\label{general_Z}
\end{equation}
where $\lambda_i$ is the admittance of the coupling along each direction, $D$ represents the dimension of the circuit, and  $\mathbf{k}=\sum_{i=1}^{D}k_i \mathbf{a}_i$ where $k_{i}=\frac{n_i\pi}{N}$ where $n_i\in\{1,2,...,2N\}$ and $i=(1,2,\dots,D)$. The asterisk sign $(^*)$ on the summation operator implies that the impedance computation must be performed considering the summation rule. For example, since the diagonally opposite corner nodes can only be defined by considering all the spatial indices $n_i$s, one must take the summation over $(n_1+n_2+\dots+n_D)\in \text{odd}$. To uniformly attach a grounding component of a single type to every node in the circuit, the admittance $z_{gnd}$ can be assigned a non-zero admittance if desired, but it will be set to 0 if not. Note that because of the translation symmetry of the periodic infinite lattice, for simplicity, we can relabel the superlattice boundaries as $n_i\in \{1,2,\dots,2N\}$ instead of $\{-(N-1),\dots,N\}$. The corner-to-corner impedance between any pair of corners in a homogeneous or heterogeneous circuit can thus be calculated using Eq.~\eqref{general_Z} by simply setting the $\lambda_i$'s in the denominator to the admittance values of the coupling along each principal direction. We will now apply this generalized expression to homogeneous and heterogeneous circuits in the following sections. We only consider passive circuit elements such as capacitors, inductors, and resistors, which can only store or dissipate energy~\footnote{But see the experiment in Ref.~\cite{zhang2022observation}, which demonstrates that such passive RLC elements can bring about non-local impedance responses in suitably designed circuits.} pumped into the circuit.

\section{Impedance results for homogeneous $RLC$ circuits with trivial unit cells in various dimensions }
In general, the corner-to-corner impedance of a homogeneous circuit constructed from passive components can be expected to increase uniformly with the circuit size because the components operate at the same phase (i.e., their admittances have the same sign). Since passive components cannot pump energy into the circuit, we intuitively expect the circuit to behave like a \textit{waveguide} as the circuit size increases. To illustrate this, we consider a 1D circuit constructed from a single type of circuit element with admittance $z_1$ and calculate the impedance between the two opposite edge nodes as new unit cells are added. Employing Eq.~\eqref{general_Z} with $D=1$, $\lambda_{1} = z_1$, $z_{gnd}=0$, and $2\cos^2(k_1/2)=1+\cos(k_1)$, the edge-to-edge impedance in 1D homogeneous circuits is obtained as
\begin{equation}
    Z_{1D}^{\text{hom}}(N)=\frac{1}{N} \sideset{}{^*} \sum_{k_1} \frac{1+\cos k_{1}}{z_{1}(1- \cos k_{1})},
    \label{Z_imp_1D_homo}
\end{equation}

\begin{figure}[t]
    \centering
    \includegraphics[width=8.6cm]{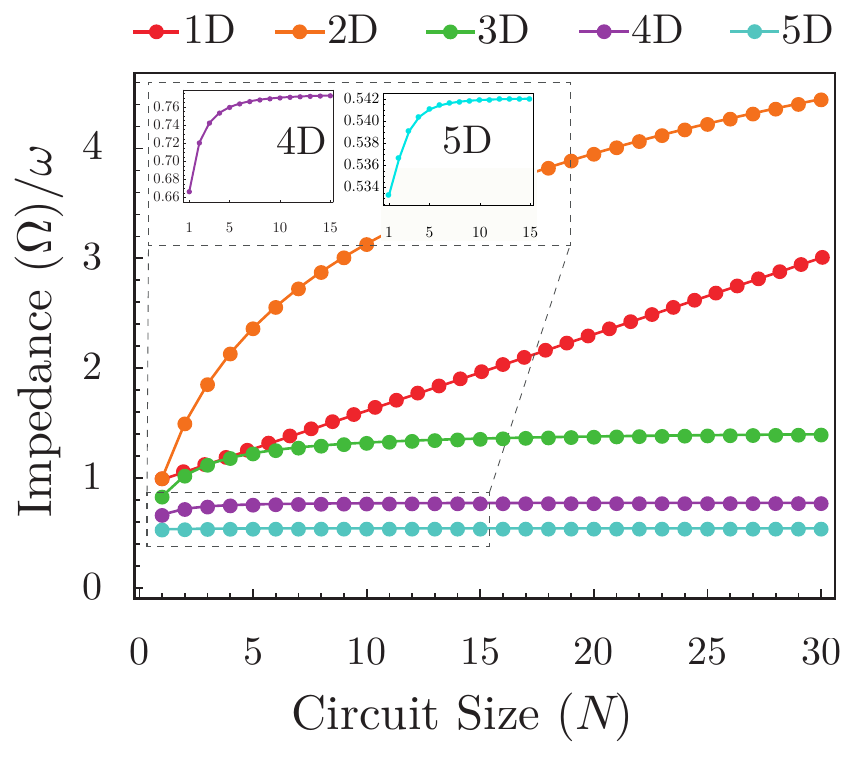}
    \caption{\textbf{Impedance across two opposite (edge) corner nodes of homogeneous (1D), 2D, 3D, 4D, and 5D circuits.} All the circuits are constructed from only a single type of capacitor $C$ and we set $C=1\ \mu \text{F}$ for all impedance computations. (The edge-to-edge impedance in the 1D chain circuit is normalized by $Z/10$ for illustration.) While the impedance between two edge nodes in a 1D chain circuit scales linearly with the size, the impedance between two opposite corner nodes scales logarithmically in 2D finite circuits and rapidly approaches a finite saturation value in three dimensions or higher, as further detailed in Appendix~\ref{appendixSaturation}.}
    \label{fig:uniform_circuits}
\end{figure}

\noindent where the asterisk means that $k_{1} = n_1 \pi/N$ and $n_1\in \{1,2,\dots,2N \}$ and $n_1\in odd$. Here, the impedance takes real values if the nodes are connected by resistors and takes imaginary values if $z_1$ corresponds to either a capacitor with an admittance of $i\omega C$ or an inductor with an admittance of $1/(i\omega L)$. Regardless of the component represented by $z_1$, the impedance increases linearly with the circuit size, as shown in Fig.~\ref{fig:uniform_circuits}. Note that Eqs.~\eqref{Imp1dsshCap} and \eqref{Imp1dsshInd} provide the same edge-to-edge impedance as Eq.~\eqref{Z_imp_1D_homo} when only one type of circuit element is present in the circuit and when $N$ is increased in steps of 2. This is because while a unit cell consists of two nodes for Eqs.~\eqref{Imp1dsshCap} and \eqref{Imp1dsshInd}, a unit cell comprises only a single node for Eq.~\eqref{Z_imp_1D_homo}.

\begin{figure}[h!]
	\centering
	\includegraphics[width=8.6cm]{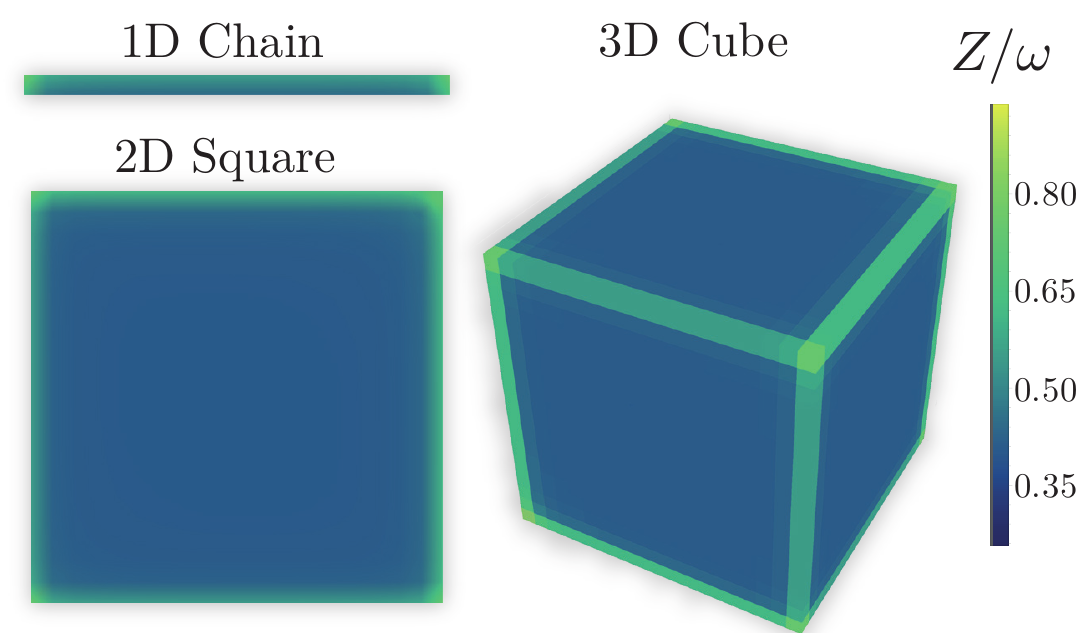}
	\caption{\textbf{Nearest-neighbor impedance \textit{distribution} in 1D, 2D, and 3D homogeneous bounded circuit arrays, with higher impedances near the boundaries.} The impedance is calculated between a node and its horizontal nearest neighbor (NN). Every cell on the density plot is colored according to the overall impedance between that node and its NN. The total impedance at a node decreases approaching the center, and increases approaching the boundaries. All the circuits are made of only a single type of capacitor with capacitance $C$ which is set as $C =1\ \mu \text{F}$.}
\label{fig:3d_cube_impedance_map}
\end{figure}

We now extend the 1D chain circuit to a 2D square circuit in which the nodes are connected by components with admittance $z_1$ along the horizontal direction and by components with admittance $z_2$ along the vertical direction. The corresponding circuit Laplacian becomes $\mathcal{L}_{2D}(k_1,k_2)=z_{1}(1- \cos k_{1}) + z_{2}(1- \cos k_{2})$. The two-point impedance for 2D circuits can be obtained by recalling Eq.~\eqref{general_Z} with $k_{1} = n_1 \pi/N$ and $k_{2} = n_2 \pi/N$ and $z_1$ and $z_2$ set to $i\omega C$ for capacitors, $ \frac{1}{i\omega L} $ for inductors, and $1/R$ for resistors. Note that $z_1$ and $z_2$ must have the same phase (i.e., both are capacitors, inductors, or resistors; or capacitors + resistors; or inductors + resistors) in order to preserve the homogeneous structure of the circuit. We can intuitively expect the impedance of the 2D homogeneous circuits to scale logarithmically with the circuit size (Fig.~\ref{fig:uniform_circuits}) because every addition of a new layer of unit cells results in a uniform increment in the overall impedance between two corner nodes with $Z=\int^N \rho \frac{d N'}{N^{'}} \sim z \log N$. Aside from the corner-to-corner impedance, the impedance between two first nearest-neighbor nodes (NNs) along the horizontal or vertical directions also scales uniformly with the size. Figure~\ref{fig:3d_cube_impedance_map} shows that the impedance between NNs is highest at the corner nodes and lower approaching the bulk nodes. This is because the current is pushed to the boundaries and accumulates there as it is reflected by the boundaries in a finite network. This trend also applies to higher dimensions such as 3D, 4D, 5D and beyond, for which interesting new phenomena can arise, and which can be feasibly implemented via circuits~\cite{bao_topoelectrical_2019,lee_electromagnetic_2018,li_emergence_2019,lee_imaging_2020,zhang_topolectrical-circuit_2020,wang_circuit_2020}.

Similar to the procedure with which we constructed the 2D homogeneous circuit, the 2D square circuit can be extended to a 3D cube circuit by simply linking every node with an additional component $z_3$ along the third direction, for which the Fourier component is written as $k_3$ where $k_3=n_3 \pi/N$. Fig.~\ref{fig:uniform_circuits} shows the two-point impedance behavior for homogeneous circuits of different dimensions as the circuit size increases. What is most remarkable is the qualitatively different behavior in higher dimensionalities~\footnote{In higher-dimensional systems where non-reciprocity also accompanies resistive non-Hermiticity, the non-local response can alter the effective dimensionality of the entire system~\cite{jiang2022dimensional,li2021quantized}.}: for homogeneous circuits is, while 1D and 2D circuits exhibit unique scaling behaviors of linear and logarithmic impedance alternations, respectively, the circuits with the dimensionality of three and above tend to saturate at some finite constant values. For instance, the two-point impedance in the 3D cube circuit starts saturating after $N\sim25$, as can be seen in Fig.~\ref{fig:uniform_circuits}. In general, for $D\geq3$, the impedance saturates to a finite value more quickly as the dimensionality of the network increases. The overall scaling in homogeneous circuits for $D\geq 3$ can be characterized by
\begin{equation}
    Z(N) \,\sim\, Z_{\text{D-dim}}^{\text{sat}} - \frac{(Z_{\text{D-dim}}^{\text{sat}})^D}{N^{D-1}},
    \label{zsaturation}
\end{equation}
\noindent where $Z_{\text{D-dim}}^{\text{sat}}$ is the saturation value, $D$ is the dimension of the circuit, and $N$ is the circuit size in terms of unit cells. To find the $Z_{\text{D-dim}}^{\text{sat}}$s, we take the limit of Eq.~\eqref{general_Z} as $N \to \infty$. We present the full analytical expression for the saturation values in Appendix~\ref{appendixSaturation}. The saturation values obtained from Eq.~\eqref{DDint} are: $Z_{\text{3-dim}}^{\text{sat}}=1.44015~\Omega$, $Z_{\text{4-dim}}^{\text{sat}}=0.774964~\Omega$, and $Z_{\text{5-dim}}^{\text{sat}}=0.542093~\Omega$, which confirm the impedance trends in the circuits for $D\geq 3$ in Fig.~\ref{fig:uniform_circuits}. For $D\geq 3$, one can analytically show that in the continuum limit, the saturation impedances can be reduced to lower-dimensional integrals, thereby expressing lower-dimensional slices of the circuit in terms of lumped effective resistances.

\section{Impedance results for heterogeneous $RLC$ circuits with nontrivial unit cells}
\subsection{2D circuits}
We now turn to heterogeneous circuits, where the corner-to-corner impedance exhibits a peculiar scaling  behavior that differs significantly from that of homogeneous circuits. Heterogeneous circuits refer to circuits that have at least two types of components for which the admittances have different complex phases, such as capacitors and inductors. To illustrate the scaling behavior, we first consider a homogeneous 1D chain circuit consisting of $N-1$ capacitors with the admittance of $z_1 = i\omega C$ along the horizontal direction $x$. We then connect $N-1$ inductors with the admittance of $z_2=1/(i\omega L)$ along the vertical direction $y$ to each node in the 1D chain circuit. This results in a two-dimensional $LC$ circuit with $N \times N$ nodes. To calculate the impedance across the diagonal corner nodes, we recall Eq.~\eqref{general_Z} and assign the admittances $z_1$ and $z_2$ to the $\lambda_i$'s such that $z_1$ and $z_2$ correspond to the principal directions. The corner-to-corner impedance of the 2D $LC$ circuit is thus given by
\begin{equation}
    Z_{2\text{D}}^{het}(N)=\frac{2}{N^2} \sideset{}{^*} \sum_{\mathbf{k}} \frac{\cos (k_1/2) \cos (k_2/2) \cos ( (k_1+k_2)/2)}{i\omega C (1- \cos k_1)+\frac{1}{i \omega L}(1- \cos k_2)},
    \label{Z_imp_2D_het}
\end{equation}
\begin{figure}[h!]
    \centering
    \includegraphics[width=8.6cm]{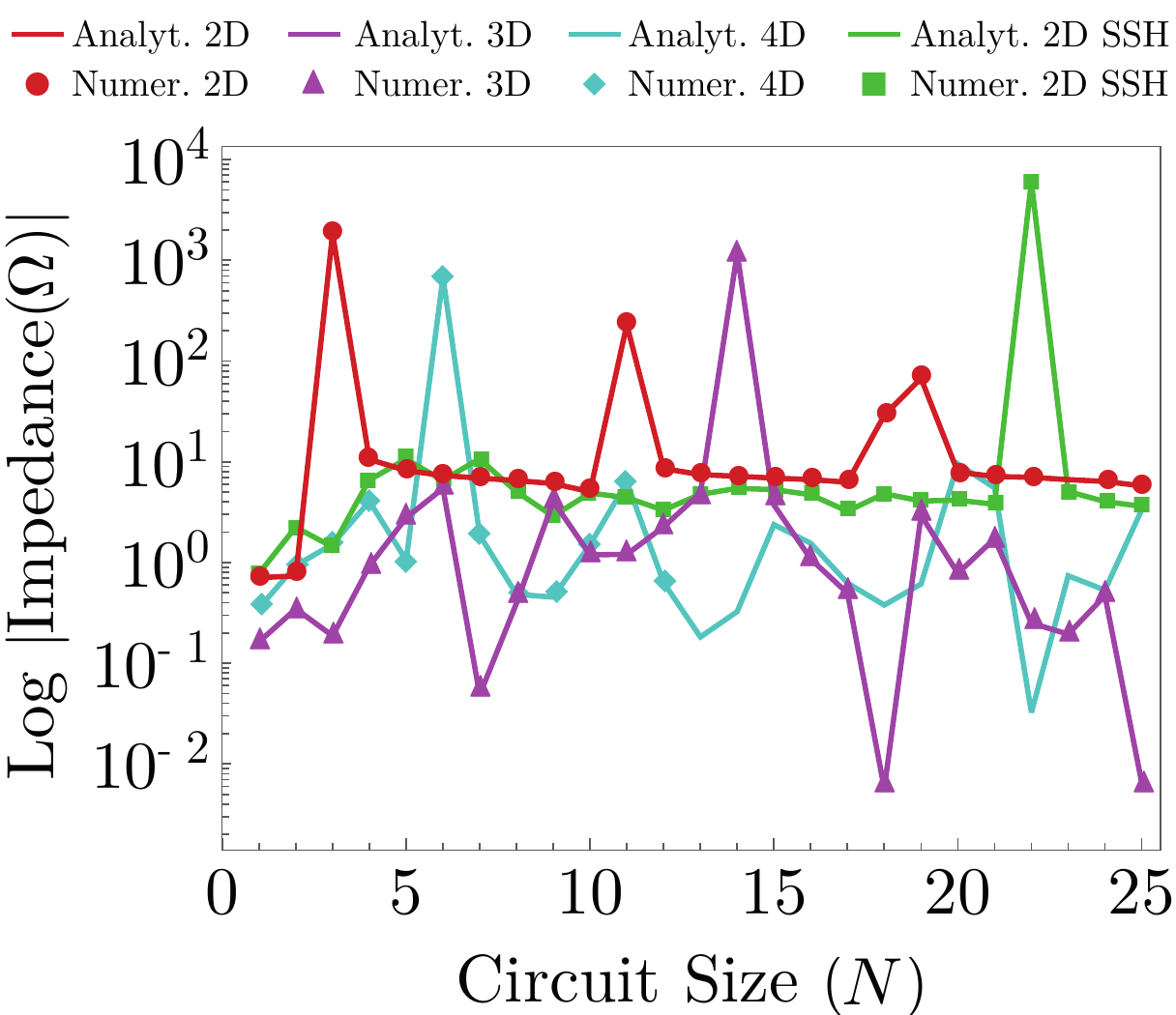}
    \caption{\textbf{Analytically and numerically calculated impedance between two opposite corner nodes in heterogeneous 2D (red), 3D (magenta), 4D (cyan), and 2D SSH (green) circuits.} Sharp peaks emerge at a certain circuit sizes in the heterogeneous circuits, unlike the logarithmic-like impedance scaling in homogeneous circuits. The component values are $\omega (C, L)=(1.7~\Omega^{-1}, 2~\Omega$) for the 2D, $\omega (C_1, C_2, L)=(2.1~\Omega^{-1}, 1~\Omega^{-1}, 2.5~\Omega)$ for the 3D,  and $\omega (C_1, C_2, L_1, L_2)=(2.5~\Omega^{-1}, 1.8~\Omega^{-1}, 2.0~\Omega, 1.0~\Omega)$ for the 4D $LC$ circuits, and  $\omega (C_1, C_2, L_1, L_2)=(2.2~\Omega^{-1}, 1~\Omega^{-1}, 2~\Omega, 1~\Omega)$ for the 2D SSH circuit.}
    \label{fig:het_circuits_imp}
\end{figure}

\noindent where the asterisk on the summation operator indicates that $\mathbf{k}=k_1 \mathbf{a_1}+k_2\mathbf{a_2}$ where $k_1 = n_1 \pi/N$ and $k_2 = n_2 \pi/N$ and $n_1$ and $n_2$ are integers satisfying $(n_1,n_2)\in \{1,2,\cdots,2N \}$ where $(n_1+n_2)$ is $odd$. This equation can be used for any corner-to-corner impedance measurement in a 2D circuit by simply considering the following summation rules ($^*$): For instance, for the impedance across two opposite corner nodes, one must take the summation over $(n_1+n_2) \in \text{odd}$, while for the impedance between the two vertical corner nodes the summation should be taken over $n_2$ is odd. Apart from the driving AC frequency $\omega$ and component parameters $C$ and $L$, which are the independent parameters in the circuit, the circuit size $N$ becomes an additional independent parameter that affects the two-point impedance. It is well known that in $LC$ resonator circuits, resonances occur at the resonant frequency $\omega = 1/\sqrt{LC}$ regardless of the value of $N$. Here, we uncover a curious phenomenon in which strong resonances occur only at particular circuit sizes. Figure~\ref{fig:het_circuits_imp} shows the occurrence of impedance jumps with the variation of the circuit size in several dimensions. The origin of these impedance resonances can be explained by considering the denominator of Eq.~\eqref{Z_imp_2D_het}. An  impedance resonance occurs when the values of $L$, $C$, and $N$ are such that there exist integers $n_1$ and $n_2$ at which the denominator of Eq.~\eqref{Z_imp_2D_het} becomes nearly zero when the terms proportional to $iC$ and $1/iL$ cancel each other almost completely. Hence, strong resonances occur only at certain circuit sizes. Moreover, the impedance peaks stem not from the numerator but arise because of the almost-vanishing denominator at particular circuit sizes in $LC$ circuits. Therefore, heterogeneous circuits with $D>2$ can potentially exhibit more interesting circuit-size dependent impedance resonances since their Laplacians have more elements than that of lower-dimensional circuits leading to more possible combinations for the cancellation in the denominator.
 
\subsection{Heterogeneous circuits in 3D and higher dimensions}
We next investigate higher-dimensional $LC$ circuits starting with a 3D cube circuit array. We construct a cube circuit by extending the 2D $LC$ square circuit along the $z$ direction using capacitors with the admittance $z_3$. The cube circuit illustrated in Fig.~\ref{network_schematics}(c) therefore comprises inductors with the admittance $1/(i\omega L)$ along the $y$ direction and capacitors with the admittances of $i \omega C_1$ and $i \omega C_2$ linking nodes along the $x$ and $z$ directions, respectively. We calculate the impedance across the corner nodes by using the $D=3$ analog of Eq.~\eqref{general_Z} and employing the Laplacian $\mathcal{L}_{3D}(k_1,k_2,k_3)=i \omega C_1(1- \cos k_{1})+1/(i \omega L)(1- \cos k_{2})+i \omega C_2(1- \cos k_{3})$ where $k_{i} = n_i \pi/N$ where $n_i\in \{1,2,\dots,2N\}$ and $i=(1,2,3)$. Due to the oddness of the discrete momentum (i.e., $(1-(-1)^{\mathbf{k}})$, refer to Eq.~\eqref{volt_final_exp}), the summation rule is determined by the corner nodes between which the impedance is to be measured. For example, to find the impedance between the diagonally opposite corner nodes $Z_{3D}^{\mathbf{1,N_a}}$, the summation is taken over $(n_1+n_2+n_3) \in odd$; while for $Z_{3D}^{\mathbf{1,N_b}}$ the rule is $(n_2+n_3) \in odd$, for $Z_{3D}^{\mathbf{1,N_c}}$ it is $(n_1+n_3) \in odd$ and finally for $Z_{3D}^{\mathbf{1,N_d}}$ it is $ n_1 \in odd$. Therefore, only the coordinate indices of the corner node opposite to the node $\mathbf{1}$ contribute to the impedance calculation and only when their sum is odd.

Similar to the 2D case, the denominator in Eq.~\eqref{general_Z} with this Laplacian leads to an anomalously large impedance measurement when there exist integer values of $(n_1,n_2,n_3) \in \{1,\cdots,2N \}$ where it becomes almost zero. The procedure we have followed to build the cube circuit from the square circuit can be extended to construct hypercube circuits. For example, a 4D $LC$ circuit can be constructed by introducing an additional direction $w$, which is perfectly feasible in 3D space due to the versatile connectivity of electrical circuits, without resorting to synthetic (non-genuine) lattice dimensions. The momentum-space Laplacian of the 4D circuit in which the nodes along the $w$ direction are linked by inductors with the admittance of $z_4=1/(i \omega L_2)$ is  
\begin{equation}
\begin{aligned}
    & \mathcal{L}_{4\text{D}}^{\text{het}}(k_1,k_2,k_3,k_4) \\&
    = i \omega \Big[ C_1(1- \cos k_{1})-\frac{1}{\omega^2 L_1}(1- \cos k_{2})\\
    & +C_2(1- \cos k_{3})-\frac{1}{\omega^2 L_2}(1- \cos k_{4}) \Big],
    \end{aligned}
    \label{Lap_4D}
\end{equation}
\noindent where $k_i = n_i\pi/N$  where $n_i \in \{1,2, \cdots, 2N\}$ and $i=(1,2,3,4)$. One can straightforwardly extend this circuit further to five dimensions (5D) or higher dimensions by invoking Eq.~\eqref{general_Z}. Whereas the impedance between two opposite corner nodes in homogeneous circuits rapidly approaches a constant saturation value, it no longer exhibits a uniform trend in the heterogeneous cube and hyper-cube circuits. Instead, large impedance peaks are observed at certain values of $N$. Figure ~\ref{fig:het_circuits_imp} shows the variation of the corner-to-corner impedance of the 4D $LC$ circuit with the circuit size $N$. %We can conclude that the violation of uniform impedance scaling occurs when the circuit Laplacian is attenuated due to the polarity the Fourier components of each direction at a certain $N$. 
Interestingly, plotting the impedance peaks due to these size-dependent resonances against the driving frequency and circuit size gives a pattern that is reminiscent of fractals, as shown in Fig.~\ref{fig:wr_N_MatrixPlot}. These fractal-like patterns are not due to the dimensionality of the circuits but rather depend on the homogeneity of circuits and, indeed, arise in the circuit size-versus-parameter diagrams of heterogeneous circuits of any dimensions, as we shall discuss in the following section.

\section{Emergent fractal-like resonances in impedance scaling behavior}
\begin{figure*}
	\centering
	\includegraphics[width=17.8cm]{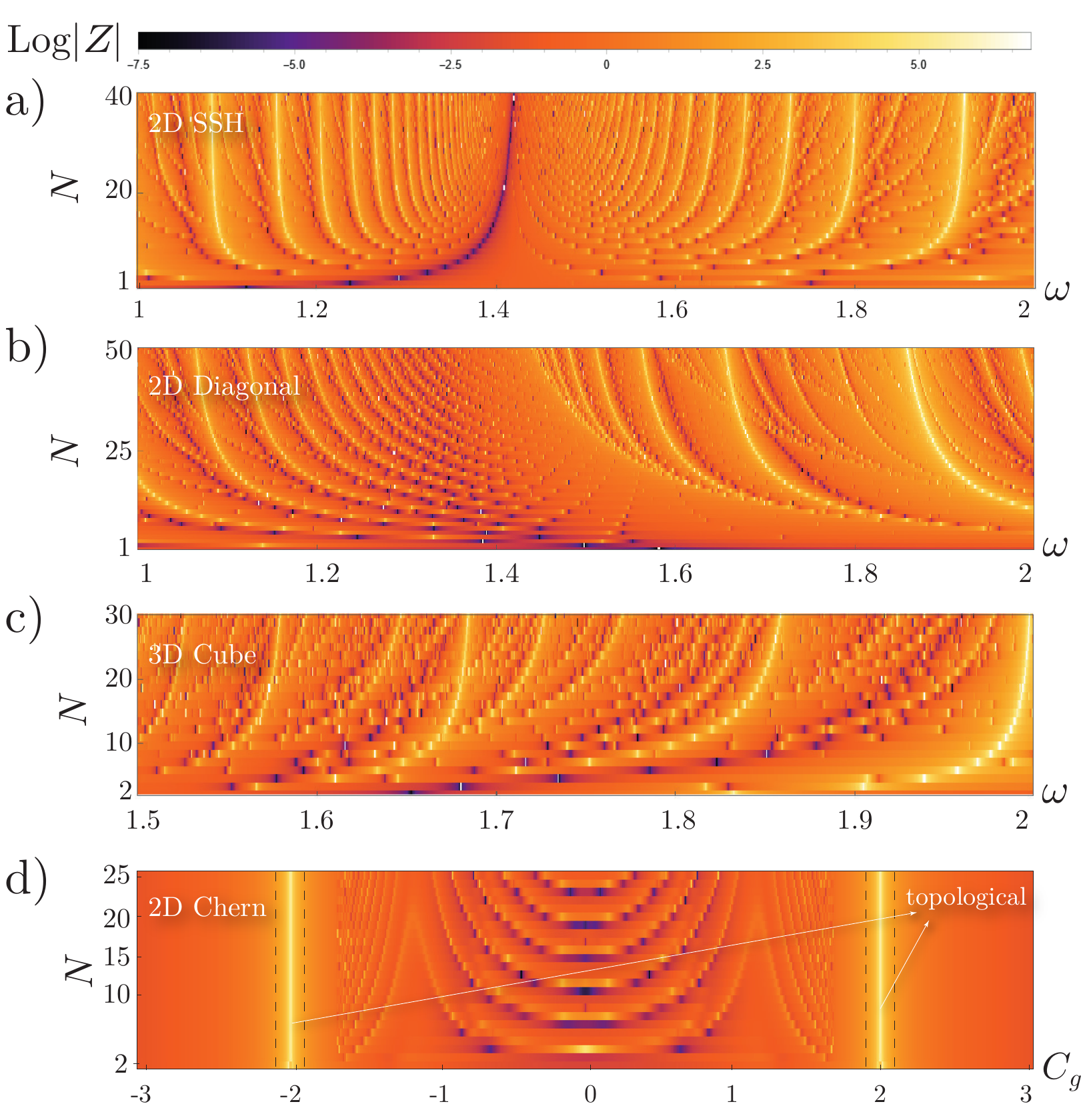}
	\caption{\textbf{Fractal-like structures emerging in the plots of the corner-to-corner impedance in heterogeneous circuits versus the circuit size $N$ and driving frequency $\omega$ [for (d), $N$ versus $C_g$]:} (a) 2D SSH circuit, (b) 2D square $LC$ circuit with diagonal cross links (obtained numerically), and (c) 3D cube circuit. The strong impedance resonances that appear as brighter branches are reminiscent of fractals. Despite the consideration of parasitic and intrinsic resistances that cause a variation of the component parameters, the structures  always survive robustly. (d) Impedance diagrams for the non-trivial topolectrical 2D Chern kink circuit. The brightest impedance branches arise in the non-trivial parameter regime and correspond to the topological zero edge modes. The circuit parameters are $(\omega C_1, \omega C_y, \omega L, R)=(1.2\ \Omega^{-1}, 0.6\ \Omega^{-1}, 0.6\ \Omega, 20\ \Omega)$.}
	\label{fig:wr_N_MatrixPlot}
\end{figure*}

Most physical systems do not change fundamentally as their system sizes are varied, except for special \textit{critical} systems exhibiting size-induced phase transitions~\cite{li_critical_2020,li_impurity_2021,rafi-ul-islam_critical_2022,siu_critical_2022}. However, in our heterogeneous circuits, we observe sharp impedance deviations from the overall trend. Whereas special resonances can be expected to occur because of specific modulations of the circuit parameters, the observed anomalous impedance resonances that arise due to the system length are unexpected. These anomalous impedance resonances can therefore be treated as violations of the logarithmic impedance scaling between two opposite corner nodes. Equation~\eqref{Z_eig} relates the two-point impedance to the eigensystem of the circuit Laplacian. The most significant difference between the eigenvalue spectrum of the Laplacians of homogeneous and heterogeneous circuits is that heterogeneous circuits typically have very small eigenvalues that can give rise to huge impedance readouts. Such a cancellation occurs in heterogeneous circuits between different components with admittance values of opposite signs, whereas no such cancellation occurs in homogeneous circuits as the admittance values have the same sign. Further, our analytical formulas show that at certain circuit sizes, there  exist combinations of the integers $n_1, n_2,\dots, n_D$ denoting the Fourier components at which the denominator of the impedance expression almost cancels out and results in strong impedance resonances. Most interestingly, the length-dependent impedance resonances seem self-organized and exhibit a fractal-like pattern in the impedance-length plots. As can be seen in Fig.~\ref{fig:wr_N_MatrixPlot}, while the strongest impedance resonances (the brightest branches) occur along certain branches, the magnitude of the impedance varies between the branches. While one can obtain higher resolution diagrams (and spanning larger parameter spaces) with more computational resources, the overall trends do not change. Since the number of bulk states is defined by the total number of nodes in the circuit, the number of branches arising in these diagrams must be the same as the number of bulk states. Therefore, every row corresponding to a selected circuit size $N$ (i.e., the impedance peaks in a row corresponding to a fixed $N$) is indeed the projection of the zero energy axis of the admittance spectrum. Consequently, because the number of admittance states increases with the circuit size $N$, the number of states intersecting with the zero-axis increases, thus resulting in the formation of curly peak lines in the diagrams. In contrast, the topological impedance resonances, as seen in Fig.~\ref{fig:wr_N_MatrixPlot}(d), form straight lines because the number of topological modes is not defined by the circuit size but rather defined by the topological invariants. As a result, the emergent fractal-like diagrams can be treated as a complete picture of the resonant properties of a circuit regardless of whether it is topological or not. We next examine two exemplary circuits to further discuss how these fractal-like diagrams differ depending on the circuit arrays.

\subsection{Two further examples for the emergent fractal-like resonances}
\subsubsection{Example: 2D square LC circuit with diagonal connections}
\begin{figure}[h!]
	\centering
	\includegraphics[width=7.5cm]{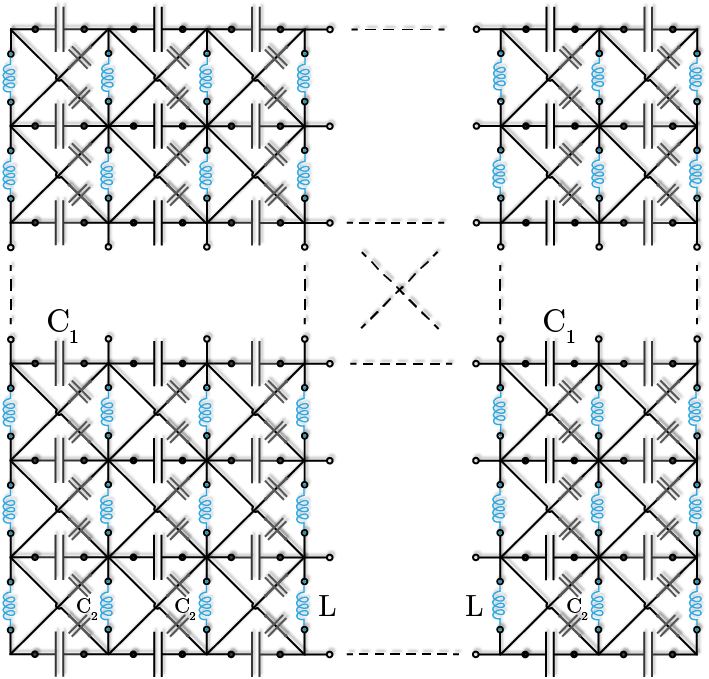}
	\caption{\textbf{2D LC circuit with diagonal cross connections with capacitors $C_2$.} The usual 2D $LC$ circuit in which the horizontal and vertical nodes are coupled with capacitors $C_1$ and inductors $L$, respectively, can be recovered by simply setting $C_2=0$. To obtain the fractal-like diagram [Fig.~\ref{fig:wr_N_MatrixPlot}(b)], the impedance is numerically measured between two opposite corner nodes as the circuit size increases.}
	\label{fig:2ddiagonalcircuit}
\end{figure}
The heterogeneous 2D $LC$ circuit that we have studied so far (see Fig.~\ref{network_schematics}(b)) can be made more complicated, which in turn leads to richer fractal-like diagrams. This circuit has capacitors $C_1$ along with the vertical direction, inductors $L$ along the horizontal direction and a second capacitor $C_2$ connecting every node with its diagonal neighbors [see Fig.~\ref{fig:2ddiagonalcircuit}]. Because of these cross connections, the circuit Laplacian has an additional term representing the nearest neighbor links with the admittance $i\omega C_2$. Therefore, the circuit Laplacian [Eq.~\ref{Z_imp_2D_het}] becomes
\begin{equation}
	\begin{aligned}
		\mathcal{L}_{2\text{D}}^{\text{cross}}(k_1,k_2) = &2 i \omega C_1 (1-\cos k_1) + \frac{2}{i \omega L} (1-\cos k_2)\\
		& + 4 i \omega C_2 (1- \cos k_1 \cos k_2)
	\end{aligned}
\end{equation}
where $k_i = n_i \pi/N$ where $n_i \in \{1,2,\dots,2N\}$ and $i=(1,2)$. As we have discussed above, the size-dependent impedance resonances are induced by the attenuation of the Laplacian. Because $G=J^{-1}$ and $V=GI$, the attenuation of the Laplacian leads to voltage accumulation at the node where the current is injected. Thus, an enormous impedance read-out ($Z \propto V$) occurs. Since the circuit Laplacian for the usual 2D $LC$ circuit has a simpler form as given in the denominator of Eq.~\eqref{Z_imp_2D_het}, it is expected that more complicated and fascinating branches may emerge in the fractal diagram of the 2D $LC$ diagonally connected circuit. As can be seen in Fig.~\ref{fig:wr_N_MatrixPlot}(b), as the circuit size increases, many resonance branches that emerge shape a pattern. Although the diagonal connections result in a well-organized fractal-like diagram, we emphasize that the diagonal links in this circuit lead to a current leakage from the physical circuit to the image circuits. In the usual 2D $LC$ circuit [see Fig.~\ref{VSpatial}(b)], the symmetrical current injection and extraction ensures that the same potential exists at the neighbor boundary nodes of the physical and image circuits so that no current leakage occurs at the boundaries. However, in a circuit with diagonal links, the diagonal links connect boundary nodes with different electrical potentials, resulting in current flow through the cross links. Therefore, it is not possible to obtain a simple analytical expression via the method of images for the diagonally connected 2D $LC$ circuit unless complicated current injection and extraction engineering is performed, which is not practical.

\subsubsection{Example: 2D Chern kink topolectrical circuit}
\begin{figure}[h!]
	\centering
	\includegraphics[width=7.65cm]{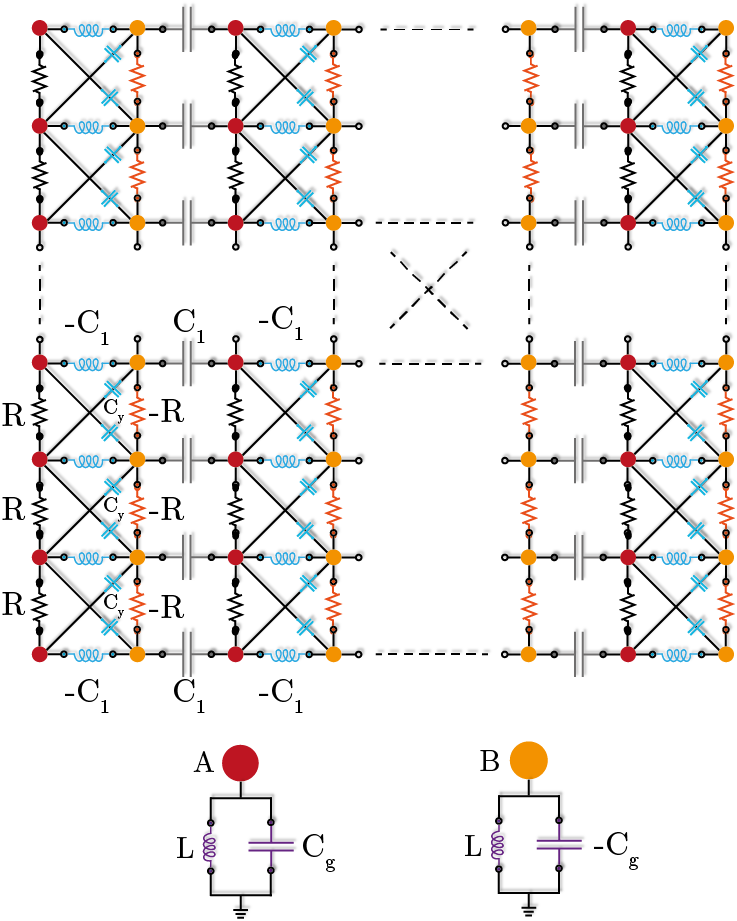}
	\caption{\textbf{The schematic of the 2D Chern kink circuit studied in Ref.~\cite{rafi2022valley}.} The circuit has two distinct nodes in a unit cell labeled as node $A$ (red circle) and node $B$ (orange circle). The two nodes within a unit cell are connected with intra-cell inductors, whose admittance is equal to that of the capacitor $-C_1$, i.e., $ L_1 = 1/ (\omega^2 C_1) = -C_1$. The horizontally adjacent unit cells are connected to each other via capacitors $C_1$. While the vertically adjacent $ A $ nodes are coupled via positive resistors $R$, two neighboring vertical $ B $ nodes are connected via negative resistors $ -R $ realized by means of negative resistance converters (NRCs). The circuit has also capacitors $C_y$ connecting $A-B$ and $B-A$ nodes along the $y$ direction. }
	\label{fig:2dvalleycircuit}
\end{figure}
Topolectrical circuits can exhibit sophisticated physical phases such as valley-dependent edge states~\cite{rafi2022valley}. For example, the valley-dependent corner or edge states can be modulated by setting the onsite capacitors $C_g$ connecting every node to the ground [refer to Fig.~\ref{fig:2dvalleycircuit}]. The negative capacitance $-C_g$ and resistance $-R$ can be achieved by using negative impedance converters (NICs) and negative resistance converters (NRCs), respectively~\cite{hofmann_chiral_2019}. The circuit Laplacian at the resonant frequency is written as
\begin{equation}
	\begin{aligned}
		\mathcal{L}_{\text{TE}}^{\text{Chern}}(k_1,k_2) = &(-C_1 + C_1 \cos k_1 + 2 C_y \cos k_2) \sigma_x \\
		& + C_1 \sin k_1 \sigma_y + (C_g + 2 R \sin k_2) \sigma_z,
	\end{aligned}
\end{equation}
where the $\sigma_\alpha$;s ($\alpha=x,y,z$) represent the Pauli matrices and where $k_i=2n_i\pi/N$ where $i=(1,2)$. As discussed in Appendix~\ref{appendix_2dssh_formula}, although a uniform grounding mechanism can be employed to isolate the topological states from the bulk states, a Chern topolectrical circuit can be designed so that it exhibits clear chiral propagating boundary states in the admittance spectrum without a uniform grounding mechanism. This is due to nontrivial center-of-mass pumping through a Laughlin-style pumping argument~\cite{vanderbilt_chapter_2006,qi_topological_2008,soluyanov_wannier_2011,lee_free-fermion_2015,liu_quantum_2016,jurgensen_chern_2022,wang2023experimental}, which has been demonstrated in a variety of classical and quantum settings~\cite{chang_experimental_2013, nash_topological_2015,ding_experimental_2019,ni_robust_2020,motruk_detecting_2020,koh_simulation_2022}. Here, due to the nearest neighbor connections and the onsite capacitors represented by the factor $\sigma_z$ in the Laplacian, the topological states are separated despite the nonuniform grounding. Therefore, this circuit is an ideal candidate to demonstrate impedance resonances arising in the fractal-like diagrams due to the topological states. For instance, the impedance resonances stemming from the edge states clearly appear in the fractal-like diagram as seen in Fig.~\ref{fig:wr_N_MatrixPlot}(d). The constant-magnitude impedance peaks can be treated as the evidence for zero-energy topological modes since the smallest eigenvalues always belong to these eigenstates. Aside from being the strongest resonances, the other remarkable property of these impedance resonances is that while the bulk states result in curved lines with the variation of the circuit size, the topological branches form distinctive straight lines. This is because even though the number of bulk states increases with the increase in circuit size, the number of topological states remains constant and the position of these states depends on the component parameters, which are fixed. Since we maintain the same component values except for the varying element ($C_g$), the topological impedance resonances appear at the same $ C_g $ value in the fractal-like diagram.

\section{Conclusion}
In conclusion, we have revealed the circuit size-dependent anomalous impedance resonances in topolectrical circuits as well as various dimensional finite $LC$ circuits. We observed that the impedance scaling in heterogeneous circuits differs from the logarithmic-like scaling exhibited by homogeneous circuits as the circuit size $N$ increases. Conventionally, frequency-dependent elements such as capacitors and inductors are considered independent circuit parameters that can be set individually. However, we have demonstrated that the strong impedance resonances in our circuits are not due to variations in these parameters, but rather to specific circuit sizes. Therefore, the circuit size becomes an independent, albeit not widely known, parameter that affects the impedance behavior of the circuit. We invoked the method of images, inspired by free-space electrostatics, as a means to calculate the corner-to-corner impedance and provided a generic exact analytical expression homogeneous or heterogeneous circuits of any dimensions. This method naturally satisfies the open boundary conditions because it ensures that the potentials of the nodes at the boundaries of the physical and image circuits are equal [see Fig.~\ref{VSpatial}(b)]. The size-dependent impedance jumps result in fractal-like patterns in the circuit size-resonant frequency plots. The existence of these patterns is common to all heterogeneous circuits but the details within the patterns are unique to each dimensionality and circuit structure. Therefore, our work establishes a framework for further investigation of anomalous impedance behaviors in more complex circuits, as well as their experimental realizations~\cite{zhang2022anomalous}.

\acknowledgments
This work is supported by the Ministry of Education (MOE) Tier-II Grant No. MOE-T2EP50121-0014 (NUS Grant No. A-8000086-01-00), and MOE Tier-I FRC grant (NUS Grant No. A-8000195-01-00). C.H.L. acknowledges support from Singapore Ministry of Education's Tier I grant (NUS Grant No. A-8000022-00-00). H.S. would like to thank the Agency for Science, Technology and Research (A*STAR) for its support of our research through the SINGA fellowship program.

\clearpage

\appendix

\section{Explicit analytical expression for the impedance of the 2D SSH circuit under periodic boundary conditions}
\label{appendix_2dssh_formula}
Here, we provide the full and explicit analytical expression for the corner-to-corner two-point impedance of the 2D PBC SSH circuit. The impedance can be obtained from the voltage difference between the lower left corner $[\mathbf{r}=(N+1,N+1),\mu=s_1]$ and upper right corner $[\mathbf{r}=(2N,2N),\mu=s_4]$ nodes given by Eqs.~\eqref{2dssh_V1} and \eqref{2dssh_V2}, respectively. There, to evaluate the voltages at the corner nodes, the Fourier transformed circuit Green's function given in Eq.~\eqref{GreenFourier} is substituted into Eqs.~\eqref{2dssh_V1} and \eqref{2dssh_V2}. Notice that when taking the inverse of the circuit Laplacian given in Eq.~\eqref{2d_ssh_lap}, any component that has zero eigenvalues should be omitted to avoid singularities. After performing the substitutions, we arrive at
\begin{equation}
	Z_{2\text{D}}^{\text{SSH}}(N) = \sum_{k_1} \sum_{k_2} \frac{\text{Numerator($k_1, k_2$)}}{\text{Denominator($k_1, k_2$)}},
\end{equation}
where the numerator and denominator are explicitly given by
\begin{widetext}
	\begin{equation}
		\begin{aligned}
			\text{Numerator} =   &\frac{ 1+(-1)^{1+n_1+n_2}}{2 i\omega^3 L_1^2 L_2^2} 
			\times \Bigg[L_1 \biggl[ 4  \omega^2 L_2 \left(2 C_1 +3 C_2\right)-4+e^{i(k_1-k_2)}\omega^2 C_1 L_2-2\omega^4 L_2^2\left(C_1^2+4 C_1 C_2+2 C_2^2\right)\\
			+& e^{-i k_1} \omega^4 C_1 C_2 L_2^2 - e^{-i(k_1-k_2)} \omega^4 C_1 C_2 L_2^2 + e^{-i k_2}\left(2-\omega^2 C_1 L_2 \right)\\
			-&e^{i k_1}\omega^2 C_1 L_2 \left(\omega^2 L_2 \left(2C_1+C_2\right)-4\right)- e^{i (k_1+k_2)} \omega^2 C_1 L_2\left( \omega^2 L_2 \left(2C_1+3C_2\right)-3\right)\\
			+&e^{i k_2} \Bigl(2+ \omega^2 L_2 \left(C_1+4C_2-2\omega^2 L_2 \left(C_1^2+2C_1 C_2+ 2C_2^2\right)\right)\Bigr)\biggr]\\
			-&4 e^{i \frac{k_1}{2}} \omega^2 L_1^2 \Bigl(2 \omega^2 L_2 \left(C_1+C_2\right)-2- \omega^4 C_1 C_2 L_2^2 \left(1- \cos(k_1)\right)\Bigr) \times \Bigl( \left( C_1+C_2 \right) \cos(k_1/2)-iC_2 \sin(k_1/2)\Bigr)\\
			+&L_2 \left(2\omega^2L_2\left(2C_2\left(1+e^{i k_2}\right)+C_1\left(1+ e^{i k_1}+e^{i k_2}+e^{i (k_1+k_2)}\right)\right) +\cos(2 k_2)+2i\sin(k_2)\left(\cos(k_2)-1\right)-1\right)\Bigg],\\
			\text{Denominator} = & \frac{4 N^2}{\omega ^4 L_1^2 L_2^2} \Bigg( - \omega^8 C_1^2 C_2^2 L_1^2 L_2^2 \Bigl(\cos \left(2 k_1\right)+3\Bigr) \\
			+&4 \omega ^4 C_1 C_2 \cos \left(k_1\right) \left(\omega ^4 C_1 C_2 L_1^2 L_2^2 -\omega ^2 L_1 L_2\left(C_1+C_2\right)  \left(L_1+L_2\right) +L_1 L_2 \cos \left(k_2\right)+L_1^2+L_2^2+L_1 L_2\right) \\
			+&4 \cos \left(k_2\right) \left(\omega ^4L_1 L_2 \left(C_1^2+C_1 C_2+C_2^2\right) -\omega ^2\left(C_1+C_2\right) \left(L_1+L_2\right) +1\right) \\
			+&4 \omega ^6C_1 C_2 L_1 L_2 \left(C_1+C_2\right) \left(L_1+L_2\right)  -4 \omega ^4 \left(C_1 C_2 L_1^2+L_1 L_2\left(C_1^2+3 C_1 C_2+C_2^2\right) +C_1 C_2 L_2^2\right) \\
			+&4\omega ^2 \left(C_1+C_2\right) \left(L_1+L_2\right)  - \cos \left(2 k_2\right) - 3 \Bigg),
		\end{aligned}
	\end{equation}
where $k_1$ and $k_2$ are the discrete momenta $k_1=n_1 \pi/N$ and $k_2=n_2 \pi/N$ where $(n_1,n_2) \in (1, 2, \cdots, 2N)$. Each $(k_1,k_2)$ contribution represents the impedance contribution from the length scale $(2\pi/k_1, 2\pi/k_2)$ in units of the lattice spacing. 
\end{widetext}

Aside from the usual impedance peaks stemming from the $LC$ resonances, a topolectrical circuit exhibits an enormous impedance readout at the resonant frequency when the circuit is topologically non-trivial. It is well known that topological systems differ from the usual bulk systems owing to their special boundary modes. These boundary states are isolated from the bulk states and appear as mid-gap states in the admittance spectrum. Such mid-gap topological states lie on the zero-energy axis, and they are associated with very small eigenvalues. Therefore, any large impedance readout at the resonant frequency can be directly related to the topological mid-gap zero-modes when they exist~\cite{rafi-ul-islam_topoelectrical_2020,rafi-ul-islam_system_2022}. For example, the topological phase is defined by the ratio of two capacitors in the usual 1D SSH circuit in Ref.~\cite{lee_topolectrical_2018} and the circuit displays a non-trivial topological phase when $C_1 / C_2 < 1$ and a trivial phase when $C_1 / C_2 > 1$. However, even though the mid-gap states are protected by topological invariants such as a non-zero integer winding number, the unequal onsite energies lead to ill-defined invariants, hence, resulting in indistinguishable topological states~\cite{gong_topological_2018,kawabata_symmetry_2019,li_winding_2015,li_winding_2015}. In the TE context, the onsite energies are represented by the diagonal elements of the circuit Laplacian. Therefore, to unveil the topological boundary states, the circuit requires uniform grounding such that all the diagonal terms in the circuit Laplacian can vanish when the driving frequency is set to the resonant frequency. Otherwise, the boundary modes join the bulk modes due to the unequal potentials at different nodes and are no longer found as mid-gap states. To uniformly ground every node in the circuit, an artificial treatment is required, which would not be possible for analytical methods. Throughout this study, since we consider an infinite periodic lattice tiled with the original physical circuits and apply the method of images to obtain the exact analytical expression for the physical circuit, it may not be possible to introduce uniform grounding into the analytical two-point impedance formula. This is because the open boundary conditions are fulfilled as a direct consequence of the method of images, which results in the Neumann boundary condition in the circuit when the circuit Laplacian has non-uniform diagonal elements. Therefore, although the non-uniform diagonal elements representing the grounding mechanism cause the disappearance of the mid-gap topological states in most cases, in some examples, the topological states can remain isolated from the bulk states and appear in fractal-like diagrams as in Fig.~\ref{fig:wr_N_MatrixPlot}d.

\section{Application of the method of images to general geometries}
\begin{figure*}[t]
	\centering
	\includegraphics[width=\textwidth]{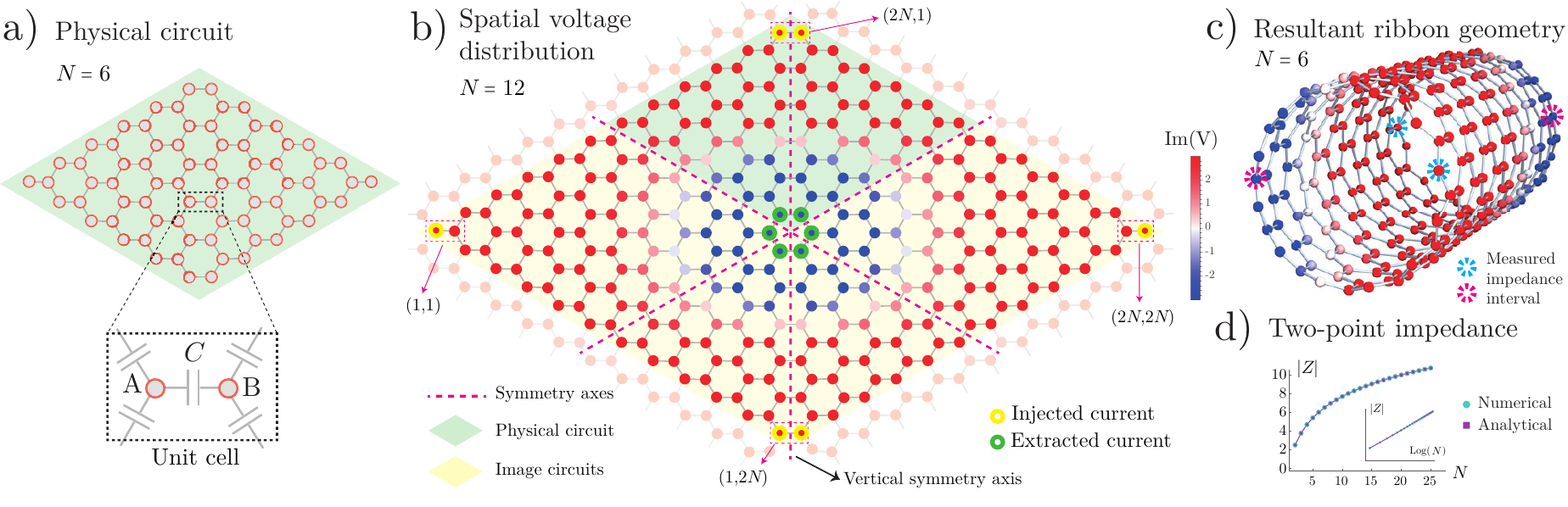}
	\caption{\textbf{An exemplary implementation of the method of images to achieve finite circuits in a general geometry using a 2D honeycomb lattice.} a) The rhombus-shaped finite physical circuit with $N=6$. A unit cell comprises two sublattice nodes $A$ and $B$, each connected by coupling capacitance $C$. b) The spatial voltage distribution response to the injected (yellow-outlined nodes) and extracted (green-outlined nodes) current configuration. The node colors represent the relative magnitude of the voltage at each node. A mirror-symmetrical potential distribution occurs about the vertical symmetry axis, ensuring the emergence of a boundary. c) The resultant distribution leads to a ribbon geometry with zigzag edges. The alternating node colors represent the voltage magnitude. d) The measured impedance between any two nodes outlined with magenta and cyan dashed circles is identical. The impedance increases logarithmically (inset) as the circuit size expands.}
	\label{fig:honeycomb}
\end{figure*}
The method of images can be applied to a general lattice model to derive an analytical expression. As discussed in the main text, the configuration of the current injection and extraction determines the spatial distribution of node voltages. To induce boundaries by utilizing the equipotential emerging on both sides of the symmetry axes, it is essential to inject and extract the current symmetrically. This results in a mirror-symmetric dispersion around the considered symmetry axes. (Note that such a symmetrical voltage distribution can only be achieved if inversion symmetry is present.) For instance, since the unit cell of the 2D SSH circuit (as well as the 2D square $LC$ circuit) exhibits translational symmetry along the $x$ and $y$ symmetry axes, the spatial potential profiles of the physical and image circuits are mirror symmetric about the $x$ and $y$ axes when there is symmetric current injection and extraction [refer to Fig.~\ref{VSpatial}(b)]. On the other hand, the symmetry axes of a hexagonal lattice provide additional opportunities for creating unique boundary designs by strategically positioning current injection and extraction sites with respect to these axes. Consequently, we can achieve diverse boundary designs by configuring the current injection and extraction points relative to the symmetry axes in a honeycomb lattice.

To demonstrate how a symmetrical current injection and extraction configuration with respect to certain symmetry axes can be utilized to establish a boundary in general geometries, we examine a 2D honeycomb lattice composed of two sublattice nodes $A$ and $B$. Our objective is to derive an analytical expression for a ribbon geometry incorporating a honeycomb lattice structure and zigzag edges. We begin by considering a geometrically rhombic honeycomb lattice with a zigzag edge design. Its momentum space circuit Laplacian reads as
\begin{equation}
	\mathcal{L}_{\varhexagon}(\mathbf{k}) = i \omega C \begin{pmatrix}
		3& -(1+e^{-i k_1}+e^{-i k_2})\\ -(1+e^{i k_1}+e^{i k_2}) & 3
	\end{pmatrix},
	\label{lap_honey}
\end{equation}
where $C$ represents the coupling capacitance, $\omega$ is the driving frequency, and $k_1$ and $k_2$ are the momentum indices. We proceed to tile the space with rhombus-shaped circuits and apply currents at specific nodes to achieve a symmetrical voltage distribution about one of the symmetry axes. Since we initially assume a periodic lattice along each direction, it is sufficient to achieve a symmetrical voltage distribution around a single axis in order to obtain the desired ribbon geometry. In Fig.~\ref{fig:honeycomb}(b), the magenta dashed lines represent the symmetry axes under consideration. Our current injection and extraction configuration results in a mirror-symmetric potential distribution about the vertical symmetry axis. This implies that the voltages of the sublattice nodes within a unit cell along the vertical axis have equal magnitudes, ensuring no current flows between these nodes. Since our circuit remains periodic in both directions, the equal voltages on either side of the vertical axis satisfy the boundary condition required to achieve a ribbon geometry.
We now move forward with deriving an analytical expression for the two-point impedance of the ribbon lattice. We recall Eq.~\eqref{volt_distr} to determine the node voltages and express the spatial positions of the nodes where the current is injected and extracted, respectively, as
\begin{equation}
\begin{aligned}
	\mathbf{\tilde{r}}_{\text{in}} \in \{ &((1,1),1),((1,2N),1),((1,2N),2),((2N,1),1),\\&((2N,1),2),((2N,2N),2)\},  \\
	\mathbf{\tilde{r}}_{\text{out}} \in \{ &((N,N),2),((N+1,N),1),((N+1,N),2),\\&((N,N+1),1),((N,N+1),2),((N+1,N+1),1)\}.
	\label{current_honey}
\end{aligned}
\end{equation}
where $((n_1,n_2),\nu)$ is shorthand for $\mathbf{r}=(n_1\mathbf{a_1}+n_2\mathbf{a_2})$ where $\mathbf{a_1}$ and $\mathbf{a_2}$ are the unit vectors, $n_1$ and $n_2$ are the position indices and $\nu\in(1,2)$ represents the sublattice nodes A and B, respectively. We can calculate the voltage at the first node as
\begin{equation}
	\begin{aligned}
		&V(\mathbf{r}=1,\mu=1)\\
		&=I\Big(G((1,1),1,(1,1),1)+G((1,1),1,(1,2N),1)\\
		&+G((1,1),1,(1,2N),2)+G((1,1),1,(2N,1),1) \\ 
		&+G((1,1),1,(2N,1),2)+G((1,1),1,(2N,2N),2) \\ 
		&-G((1,1),1,(N,N),2)-G((1,1),1,(N+1,N),1) \\ 
		&-G((1,1),1,(N+1,N),2)-G((1,1),1,(N,N+1),1) \\ 
		&-G((1,1),1,(N,N+1),2)-G((1,1),1,(N+1,N+1),1)\Big).
	\end{aligned}
	\label{honeyVoltage}
\end{equation}
By employing the momentum space Green's function in Eq.~\eqref{GreenFourier} and $\mathcal{L}_{\varhexagon}^{-1}(\mathbf{k})=G(\mathbf{k})$, the voltage at node $V(\mathbf{r}=1,\mu=1)$ is obtained as
\begin{widetext}
	\begin{align}
		V(\mathbf{r}&=1,\mu=1) = \frac{1}{4i\omega C N^2(3-\cos(k_1) - \cos(k_2) -\cos (k_1-k_2))}\\
		&\times \left( 5(1+\cos(k_1)+\cos(k_2))+2\cos(k_1-k_2)+\cos(k_1+k_2)-\cos(\frac{N-1}{N}(n_1+n_2)\pi) \right.\\
		&\left. -2\cos(\frac{N-1}{N}(n_1-n_2)\pi)-4\cos(\frac{N-1}{N}n_1\pi-n_2\pi)-\cos(\frac{N-1}{N}n_1\pi+n_2\pi)-\cos(\frac{N-1}{N}n_2\pi+n_1\pi)\right),
	\end{align}
\end{widetext}
where $k_1=n_1\pi/N$ and $k_2=n_2\pi/N$. To simplify the above equation, we assume that $n_1\in(1,2,3,\cdots,2N)$ and $n_2\in(1,3,5,7,\cdots,2N-1)$. Owing to the translation symmetry in our circuit, the impedance between node 1 and its corresponding counterpart located at the opposite edge [i.e., $(N,N)$] is given by $Z_{\varhexagon}(N) = 2V(1, 1) = 2V(N, N)$. The impedance between any pair of nodes denoted by magenta and cyan dashed circles in Fig.~\ref{fig:honeycomb}(c) is identical, a consequence of the symmetrical voltage distribution. In Fig.~\ref{fig:honeycomb}(d), we plot our numerical and analytical impedance calculation results for the ribbon circuit depicted in Fig.~\ref{fig:honeycomb}(c). There is an exact correspondence between the two sets of results. Our results also show that the logarithmic dependence of the circuit impedance on the circuit size $N$ applies for two-dimensional circuits regardless of the lattice geometry [see Fig.~\ref{fig:honeycomb}(d)].

\section{Saturation in homogeneous circuits}\label{appendixSaturation}
To find the finite saturation values in homogeneous circuits when $D>2$, let us first consider a homogeneous 3D cube circuit constructed by a typical resistor with the resistance $R$ in each principal direction. Now, it is essential to determine the electric potential at the opposite corner nodes whereby the corner-to-corner impedance as a function of the circuit size can be simply found by $Z=2|V|/I=2G$ due to the translation symmetry implying that $V(\mathbf{r}) = - V(-\mathbf{r})$. To impose the boundary conditions, symmetrical current injection and extraction [for example, from Fig.~\ref{network_schematics}(c)] are performed such that the impedance between two opposite corner nodes in terms of the voltage distribution is written as
\begin{equation}
\begin{aligned}
	Z_{3\text{D}}=2\Big( & G(0,0,0)+G(1,0,0)+G(0,1,0)+G(0,0,1)\\
	&+G(1,1,0)+G(1,0,1)+G(0,1,1)+G(1,1,1) \\ 
	&-G(N,N,N)-G(N+1,N,N)\\ 
	&-G(N,N+1,N)-G(N,N,N+1) \\ 
	&-G(N+1,N+1,N)-G(N+1,N,N+1) \\ 
	&-G(N,N+1,N+1)- G(N+1,N+1,N+1)\Big),
\end{aligned}
\label{cubeV1}
\end{equation}
where $G(n_1,n_2,n_3)$ is the short notation of $G(\mathbf{r})$ [e.g., $G(\mathbf{r}) \rightarrow G(0,0,0)$ or $G(\mathbf{r}+N\mathbf{u}+\mathbf{a_1}) \rightarrow G(N+1,N,N)$], where $\mathbf{r} =\sum_i^{D=3} n_i \mathbf{a_i}$ where $n_i$s are the integers varying from 1 to $2N$ and where $\mathbf{u} = \mathbf{a_1} + \mathbf{a_2} + \mathbf{a_3}$ being the unit vector. As accomplished in Sec.~\ref{secRLC}, the circuit Green's function given by Eq.~\eqref{Green_r} is inserted into the above equation. Because the circuit nodes are connected by resistors with the admittance $1/R$, the corresponding circuit Laplacian is written as $\mathcal{L}_{3D}^{\text{sat}}(\mathbf{k})= (2/R)(3-\cos k_1 - \cos k_2 - \cos k_3)$ where $\mathbf{k}=\sum_i^{D=3} k_i \mathbf{a_i} $ and where $k_{i}=n_i\pi/N$ where $n_i\in \{1,2,\dots,2N\}$ and $i=(1,2,3)$. However, for the sake of simplicity, we henceforth set $R=1\Omega$ in the derivation. By inserting the corresponding Laplacian substituted for $G(\mathbf{k})^{-1}$, we arrive
\begin{equation}
	\begin{aligned}
		&Z_{3\text{D}}(N)=\frac{2}{(2N)^3} \sum_{n_1=1}^{2N}\sum_{n_2=1}^{2N}\sum_{n_3=1}^{2N}\big( 1-e^{i\pi(n_1+n_2+n_3)}\big)  \\ 
		&\  \ \times \frac{8\cos\left(\frac{n_1\pi}{2N}\right) \cos\left(\frac{n_2\pi}{2N}\right) \cos\left(\frac{n_3\pi}{2N}\right) \cos\left(\frac{(n_1+n_2+n_3)\pi}{2N}\right)}{2\left(3-\cos\left(\frac{n_1\pi}{N}\right) - \cos\left(\frac{n_2\pi}{N}\right) - \cos\left(\frac{n_3\pi}{N}\right)\right)}.
	\end{aligned}
\label{3Dimp}
\end{equation}
Here, we obtain the analytical formula for the corner-to-corner impedance in the 3D homogeneous cube circuit. We can now generalize the corner-to-corner impedance in $D$ dimensions and write
\begin{equation}
	\begin{aligned}
		Z_\text{D-dim}(N)=&\frac{1}{N^D} \sum_{n_1=1}^{2N}...\sum_{n_D=1}^{2N}\big( 1-e^{i\pi\sum_{ i}^Dn_i}\big)  \\ 
		&\times \frac{\cos\left( \sum_{ i}^D\frac{n_i\pi}{2N}\right) \times \prod_{i}^D\cos\left(\frac{n_i\pi}{2N}\right) }{D-\sum_{ i}^D\cos\left( \frac{n_i\pi}{N}\right) }.
	\end{aligned}
	\label{DDimp}
\end{equation}
Here, $D$ refers to the circuit dimension, $n_i$s are the Fourier components varying between $1$ and $2N$ along each direction, and $N$ is the circuit size. Notice that because we have set $R=1\Omega$ earlier, there is no term representing the unit admittance. However, one can multiply the denominator of Eq.~\eqref{DDimp} by a unit admittance if required because the admittance is a common factor in the Laplacian since the circuit nodes are linked by single-type components. The impedance between two opposite corner nodes calculated through the above expression saturates as $N$ approaches the continuum limit, i.e., $Z_{\text{D-dim}}^{\text{sat}}\approx\lim_{N \to\infty}Z_\text{D-dim}(N)$. The presence of saturation impedances in the large-$N$ limit in dimensions $D\geq 3$ suggests a convergent integral expression for the corner-to-corner impedances in this regime. In the large-$N$ limit, the evenness and oddness of the indices should have negligible effects if the sum can be approximated by an integral, since that entails shifts of $\pi/N$ to the momenta. Noting that the factor $1-e^{i\pi(\sum_{i}^D n_i)}$ effectively eliminates half of the terms, we have the approximation
\begin{equation}
	\begin{aligned}
		Z_\text{D-dim}(N\rightarrow\infty)\approx& \frac1{\pi^D}\int\displaylimits_{[0,2\pi]^D}\frac{\cos\left(\sum_i^D \frac{k_i}{2}\right) \times \prod_i^D\cos\frac{k_i}{2}}{D-\sum_i^D \cos k_i}\,d^D\mathbf{k}
	\end{aligned}
	\label{DDint}
\end{equation}
for $D\geq 3$. In practice, a very small positive term may have to be added to the dominator to make the integral converge; physically, this corresponds to inevitable parasitic resistances. For instance, we obtain $		Z_\text{3-dim}(N\rightarrow\infty)=1.44015~\Omega$, $Z_\text{4-dim}(N\rightarrow\infty)=0.774964~\Omega$ and $Z_\text{5-dim}(N\rightarrow\infty)=0.542093~\Omega$, which are very close to the values of $Z_{\text{3-dim}}(300)=1.43583~\Omega$, $Z_{\text{4-dim}}(50)=0.774739~\Omega$ and $Z_{\text{5-dim}}(40)=0.542011~\Omega$ computed from Eq.~\eqref{DDimp}.

\subsubsection{Dimensional reduction of impedance formula}
Interestingly, the corner-to-corner impedance can be dimensionally reduced to a momentum integral in a lower number of dimensions, albeit with a seemingly more sophisticated integrand. To start, we separate out the last momentum coordinate $k_D=k$ by writing the integrand $\frac{\cos\left(\sum_i^D \frac{k_i}{2}\right)\prod_i^D\cos\frac{k_i}{2}}{D-\sum_i^D \cos k_i}$ of Eq.~\eqref{DDint} as $\left(\prod\limits_i^{D-1}\cos\frac{k_i}{2}\right)\frac{\cos \frac{k}{2}\cos\left(a+\frac{k}{2}\right)}{D-b-\cos k}$ where $a=\sum\limits_i^{D-1} \frac{k_i}{2}$ and $b=\sum\limits_i^{D-1} \cos k_i$. The $k$-dependent fraction on the right can be integrated as follows: $\int_0^{2\pi} \frac{\cos \frac{k}{2}\cos\left(a+\frac{k}{2}\right)}{D-b-\cos k}dk =\pi\left(\sqrt{\frac{D-b+1}{D-b-1}}-1\right)\cos a $. Since $\int_{[0,2\pi]^{D'}}\cos\left(\sum\limits_i^{D'} \frac{k_i}{2}\right)\prod\limits_i^{D'}\cos\frac{k_i}{2}\,d^D\bold{k}=\pi^{D'}$ for any $D'$ (Here and below, we denote $D'=D-1$ for brevity), we obtain $Z_\text{D-dim}(N\rightarrow\infty)\approx $
\begin{equation}
	\begin{aligned}
		-1+\int\displaylimits_{[0,2\pi]^{D'}} \frac{\cos\left(\sum\limits_i^{D'} \frac{k_i}{2}\right)\prod\limits_i^{D'}\cos\frac{k_i}{2}}{\pi^{D'}}\sqrt{\frac{D+1-\sum\limits_i^{D'} \cos k_i}{D-1-\sum\limits_i^{D'} \cos k_i}}\,d^{D'}\bold k.
	\end{aligned}
	\label{DDint2}
\end{equation}
In general, we can continue with this dimensional reduction procedure to obtain $	Z_\text{D-dim}(N\rightarrow\infty)\approx$
\begin{equation}
	\begin{aligned}
		\int\displaylimits_{[0,2\pi]^{D-d}} \frac{\cos\left(\sum\limits_i^{D-d} \frac{k_i}{2}\right)\prod\limits_i^{D-d}\cos\frac{k_i}{2}}{\pi^{D-d}}f_d\left(D-d-\sum\limits_i^{D-d}\cos k_i\right)\,d^{D-d}\bold k
	\end{aligned}
	\label{DDintred}
\end{equation}
where $d$ is the number of reduced dimensions and $f_d(x)$ is a weightage function that encapsulates the effects of dimensional reduction:
\begin{widetext}
	\begin{align}
		f_0(x) &= \frac1{x}\\
		f_1(x) &= \sqrt{1+\frac{2}{x}}-1\\
		f_2(x) &= \frac{\left(x^2+5 x+8\right) K\left(-\frac{4}{x^2+4 x}\right)-x (x+4) E\left(-\frac{4}{x^2+4 x}\right)-2 i K\left(\frac{(x+2)^2}{x (x+4)}\right)}{\pi\sqrt{x (x+4)}}+\frac1{\pi}K\left(\frac{4}{(x+2)^2}\right)+\frac{2 i K\left(\frac{x (x+4)}{(x+2)^2}\right)}{\pi(x+2)}-1,
	\end{align}
\end{widetext}
where $E(y)=\int_0^{\pi/2}\sqrt{1-y\sin^2\theta}\,d\theta$ and $K(y)=\int_0^{\pi/2}1/\sqrt{1-y\sin^2\theta}\,d\theta$ are elliptic integrals. Further expressions of $f_d$, $d>2$ exist in principle, although they would be much more complicated. Note that $f_0$ is just the inverse Laplacian spectrum for the unbounded circuit; the other terms in the integrand keeps track of the source/sink as well as boundary effects, as we have obtained from the method of images. As an illustration,
\begin{equation}
	Z_\text{3-dim}(N\rightarrow\infty)\approx \frac1{\pi}\int_0^{2\pi}\cos^2\frac{k}{2}\,f_2(1-\cos k)dk.
\end{equation}

\bibliography{impedance_arxiv_v3}
\end{document}